\documentclass[aps, prd, longbibliography, 11pt,superscriptaddress,
notitlepage]{revtex4-1} 
\usepackage{changepage}
\usepackage{stix} 
\usepackage[export]{adjustbox}
\usepackage{tikz-cd, mathtools} 
 \usetikzlibrary{calc}
 \usetikzlibrary{arrows}
\usetikzlibrary{graphs,decorations.pathmorphing,decorations.markings}
 \usepackage{mathrsfs}
 \usepackage[cal=boondoxo]
 {mathalpha}
\usepackage{lipsum}
\usepackage[utf8]{inputenc} 
\usepackage{graphicx}
\usepackage{comment}
\usepackage{bbding}
\usepackage{amsmath}
\usepackage{xcolor}
\usepackage{verbatim}
\usepackage[]{algorithm2e}
\usepackage{anyfontsize}

\usepackage[T1]{fontenc}

\newcommand{\optional}[1]{}

\newcommand{\rank}{\textrm{rank\,}}

\newcommand{\Iall}{\mathcal{I}}
\newcommand{\I}[1]{\mathcal{I}(#1)}
\newcommand{\sumiall}{i\,\in\,\Iall}

\newcommand{\smallolistsymbol}{\,
  \tikz[baseline=-.6ex, line width=.8pt]{
   \draw (0,0) circle (3.8pt);
    \draw (-0.2em, 0.12em) -- (0.2em, 0.12em);
    \draw (-0.2em, 0em)     -- (0.2em, 0em);
    \draw (-0.2em, -0.12em) -- (0.2em, -0.12em);
  }\,
}

\newcommand{\bigolistsymbol}{\,
  \tikz[baseline=-.6ex, line width=1.45pt]{
   \draw (0,0) circle (7pt);
    \draw (-0.35em, 0.21em) -- (0.35em, 0.21em);
    \draw (-0.35em, 0em)     -- (0.35em, 0em);
    \draw (-0.35em, -0.21em) -- (0.35em, -0.21em);
  }\,
}

\newcommand{\smallbullet}{\mathbin{\vcenter{\hbox{\scalebox{0.45}{$\bullet$}}}}}

\newcommand{\bigolist}{\mathop{\bigolistsymbol}\limits}

\newcommand{\olist}
{\mathop{\smallolistsymbol}\limits}

\newcommand{\bigsoldersymbol}{\,
  \tikz[baseline=-.6ex, line width=1.45pt]{
    \draw (0,0) circle (7pt);
    \draw (90:7pt) -- (270:0pt);
    \draw (210:7pt) -- (30:0pt);
    \draw (330:7pt) -- (150:0pt);
  }\,
}

\newcommand{\bigsolder}{\mathop{\bigsoldersymbol}\limits}

\newcommand{\solder}{\,
  \tikz[baseline=-.75ex, line width=0.75pt]{
    \draw (0,0) circle (3.6pt);
    \draw (90:3.6pt) -- (270:0pt);
    \draw (210:3.6pt) -- (30:0pt);
    \draw (330:3.6pt) -- (150:0pt);
  }\,
}

\newcommand{\subscriptsolder}{
  \tikz[baseline=-.4ex, line width=0.37pt]{
    \draw (0,0) circle (1.8pt);
    \draw (90:1.8pt) -- (270:0pt);
    \draw (210:1.8pt) -- (30:0pt);
    \draw (330:1.8pt) -- (150:0pt);
  }
}

\newcommand{\postdot}[1]{#1^{\smallbullet}}

\makeatletter

\newcommand*\notocchapter[1]{%
  \if@openright\cleardoublepage\else\clearpage\fi
  \thispagestyle{empty}\global\@topnum\z@
  \@afterindenttrue
  \let\@secnumber\@empty
  \@makeschapterhead{#1}\@afterheading
}

\makeatother

\newcommand{\nocontentsline}[3]{}
\newcommand{\tocless}[2]{\bgroup\let\addcontentsline=\nocontentsline#1{#2}\egroup}

\usepackage{hyperref}
\begin{document}
\title{High-resolution decomposition of canonical Hamiltonian systems}
\author{Kaja Krhac}
\affiliation{School of Mathematics and Natural Sciences,
University of Wuppertal
Gaußstraße 20, 42119 Wuppertal, Germany}
\author{Frederic P. Schuller}
\email{Corresponding author: f.p.schuller@utwente.nl}
\affiliation{Department of Applied Mathematics,
University of Twente,
P.O. Box 217, 7500 AE Enschede, The Netherlands}
\begin{abstract}
  \noindent We identify the topological and dynamical conditions for a Hamiltonian system to possess port-Hamiltonian reformulations with increasing resolution of structural detail that remains hidden in the Hamiltonian theory. We find that the principally attainable resolution grows exponentially with the number of factor manifolds into which the configuration space of a canonical Hamiltonian system decomposes and is tempered only by degeneracies of the original Hamiltonian.   
\end{abstract}

\maketitle
\vspace{-.6cm}
\begingroup
  \makeatletter
\setlength{\baselineskip}{14pt} 
  \setlength{\parskip}{3pt}
  \tableofcontents
  \makeatother
\endgroup



\newpage
\section{Introduction}\label{sec:intro}
\noindent 
Geometric Hamiltonian theory provides a rigorous and powerful framework for classical physics. But two intuitive concepts, which we routinely think and talk about when using Hamiltonian theory, are not reflected at all in its mathematical formalism: subsystems and the energy flow between them due to their interaction. The wording of Newton's third axiom, for instance, cannot be adequately expressed in Hamiltonian theory. Its implementation is effectively relegated to the modelling process without any further structural insight. Even the structural hallmark of Hamiltonian theory, overall energy conservation, merely allows one to think intuitively about formally undefined energy flows between formally equally undefined subsystems. 

But this lack of a formalism that captures these standard ideas prevents one from pursuing them with any mathematical machinery and thus obstructs one from taking them far beyond what one's intuition immediately suggests. Also, based on intuition alone, even the most basic questions remain unclear: How many subsystems does an interacting $N$-particle system consist of? We find it is up to $2^N-1$ subsystems. How are these subsystems and the energy flows between them defined on the phase space of the canonical Hamiltonian system? 
We find that the phase space cannot serve this purpose at all and that one must elevate the geometry to a carefully constructed bundle over the mere phase space.

These two clear answers to the two ill-defined questions above obviously employ particular formal definitions beyond geometric Hamiltonian theory. In this article, we indeed build on the central concepts of port-Hamiltonian theory~\cite{maschke1992port, vanderschaft2014introductory}, which provides a rigorous geometric description of open systems and their composition to either still open or finally closed systems across different physical domains~\cite{vanderschaft2006}. This modular structure has earned the theory a celebrated role in high-end engineering whose bread and butter business is to \emph{compose} various physical systems \cite{rashad2020,duindam2009} to ingenious gadgets Nature did not come up with. 

In this article, in contrast, we employ port-Hamiltonian concepts to systematically \emph{decompose} canonical Hamiltonian systems. The two answers above are, in fact, simply the answers one immediately obtains when one maps a canonical Hamiltonian system with geometric rigour to a high-resolution port-Hamiltonian description and then asks the respective questions within that formal framework. Our main result is indeed the identification of the precise mapping that takes one from Hamiltonian to port-Hamiltonian theory. We find that this mapping is purely topological, but otherwise serves as the direct conceptual analogue of the Legendre transformation that takes one to Lagrangian theory instead. Just as Lagrangian theory is in many ways more versatile than Hamiltonian theory, since it effortlessly accommodates gauge theories, non-holonomic constraints and dissipative systems~\cite{abraham1978foundations}, port-Hamiltonian theory is vastly more versatile than Hamiltonian theory. It is thus immensely practical to now also have, with the main result of this paper, a canonical way of transition to port-Hamiltonian theory, even beyond the conceptual clarifications the latter affords. 

\newpage
\noindent The technical reasoning of this article starts with the observation that every canonical Hamiltonian system can  be rewritten trivially in terms of the key entities of the port-Hamiltonian formalism, i.e., an  energy storage and a Dirac structure~\cite{vanderschaft1998affine, yoshimura2006dirac1, vanderschaft2014introductory}. This trivial rewriting is always possible but does not reveal any new detail yet that would not already be on display in the Hamiltonian formulation.  

The necessary condition to achieve a genuinely higher resolution of detail is a \emph{topological} one, namely that the configuration space of the Hamiltonian system decomposes into a product
\begin{equation}\label{introfac}
    M = Q_1 \times \dots \times Q_N
\end{equation}
of more than one lower-dimensional smooth manifold. If the configuration space is a 2-torus, for instance, this is of course possible and unique since $T^2=S^1 \times S^1$ decomposes the torus into a product of circles, but if the configuration space is a 2-sphere, there is no factorisation and hence no higher-resolution port-Hamiltonian reformulation. In case the  configuration manifold is additionally equipped with a Riemannian structure $g$, as routinely would be the case in order to provide sufficient structure for the formulation of kinetic terms in the Hamiltonian function, it is well-known \cite{deRham1952, EschenburgHeintze1998} that such decompositions become more rigid and finally unique for complete  Riemannian configuration spaces if one proceeds down to irreducible factors, except for a possible Euclidean factor which can be decomposed in infinitely many different ways.

Through modest extension and exploitation of canonical constructions on product manifolds \cite{yoshimura2006dirac1,yoshimura2006dirac2}, 
we show that any factorisation (\ref{introfac}) of the configuration manifold, irrespective of uniqueness or complete irreducibility of its factors, effects a generic factor of $2^N-1$ in structural resolution. The detail that emerges this way indeed comes in the form of a network of up to $2^N-1$ energy storages that are interconnected by a purely topologically determined routing structure for energy flows between them.
This principally exponential increase in resolution is only tamed 
 if one starts from an insufficiently generic, but otherwise perfectly fine Hamiltonian, which causes some energy storages to effectively vanish. Sufficient genericity of the Hamiltonian thus constitutes the subordinate \emph{dynamical} condition to obtain formulations of a specific higher resolution. 
 In short, the configuration space topology determines the maximally attainable resolution, while the degeneracies of the  Hamiltonian determine by how much this maximal resolution is reduced.

Our systematic geometric and differential topological  construction --- of the high-resolution energy storages and of the high-resolution routing structure that interconnects them --- is built on splitting and merging operations of spaces, structures on these spaces and a subtle use of bundles with overlapping base spaces. But once the dust settles, the resulting port-Hamiltonian formulation   ultimately reveals a simple recipe for how to \emph{directly} obtain the higher-resolution formulation of a canonical Hamiltonian system along a known factorisation of its configuration space, without  having to retrace the constructive steps taken in this article each time:
\newpage
\noindent 
We indeed find that two families of maps, 
\begin{equation}\label{musigmapath}
    \begin{tikzcd}[row sep=large, column sep=large]
    T^*Q_n 
    \arrow[r,"{\mu_n}"]
    &  
    T^*(Q_1\times\dots\times Q_N)
    \arrow[r, "{\sigma_{i_1\dots i_r}}"]
    &  T^*Q_{i_1} \times\dots\times T^*Q_{i_r}\,,
    \end{tikzcd}
\end{equation}
with an index $n$ ranging over $1, \dots, N$ and a multi-index $i_1\dots i_r$ ranging over the set $\Iall$ defined at the end of the introduction, single-handedly effect the transition from the Hamiltonian to the high-resolution port-Hamiltonian formulation associated with the configuration space factorisation (\ref{introfac}). 
The explicit definition of these maps in  terms of canonical splitting and merging maps along factorisations is ultimately identified in (\ref{musigmadef}). Together, these two maps effect the complete transformation 
\begin{center}
\begin{tikzpicture}
    \draw (0, 0) node[inner sep=0] {\includegraphics[width=.47\linewidth]{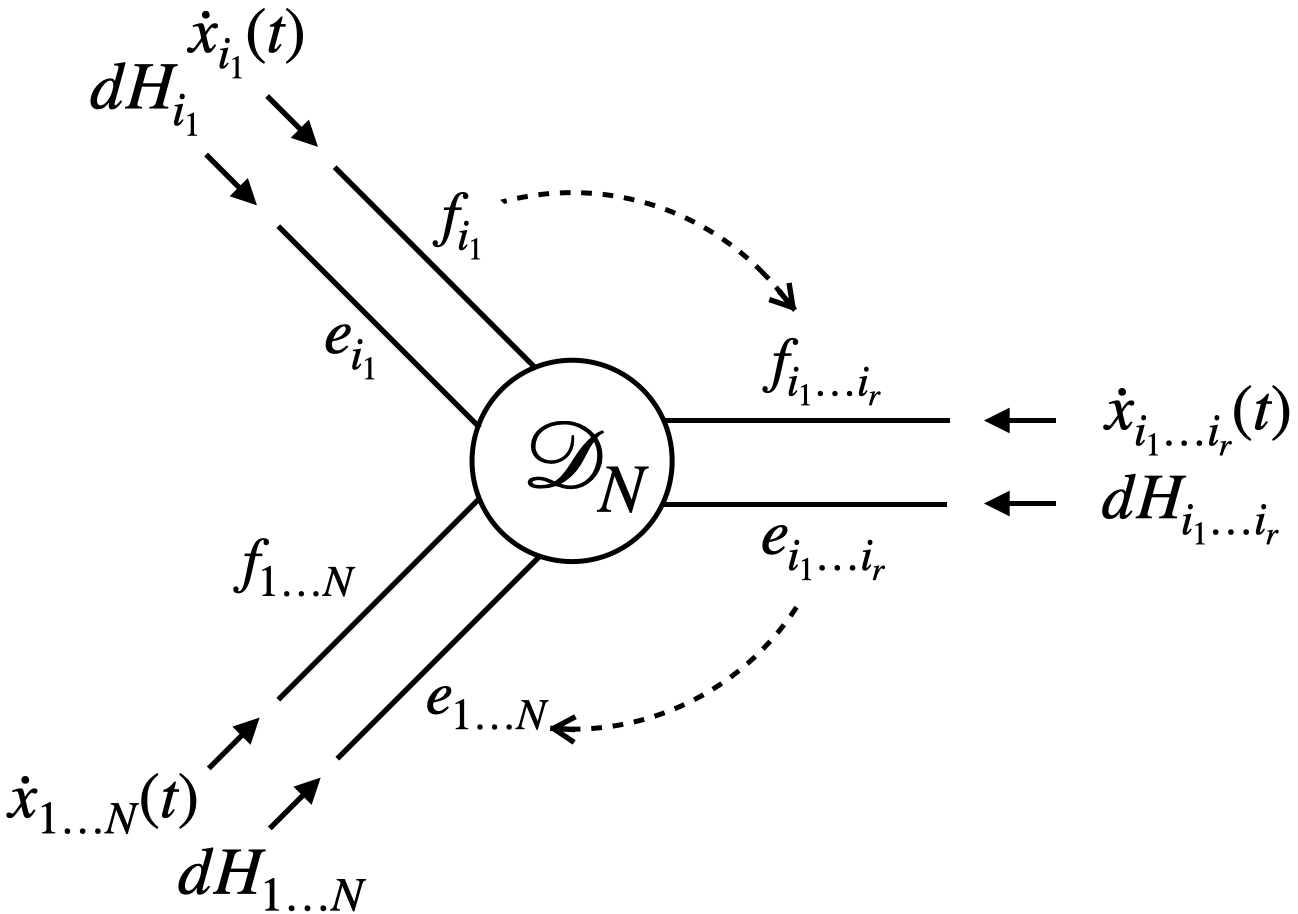}};
    \draw (-8, 0) node {$\Omega_\flat(\dot x(t)) =dH_{x(t)}$};
    \draw[line width=1pt] (-5.5,1mm) -- (-5.5,-1mm);
    \draw[line width=1pt, ->] (-5.5,0) -- (-4.2,0);
\end{tikzpicture}
\end{center}
\vspace{-12pt}
of a canonical Hamiltonian system formulated on $T^*M$ to the corresponding  highest-resolution port-Hamiltonian system on $T^*Q_1\times\dots\times T^*Q_N$ by providing, according to precise conditions formulated in section \ref{sec:hiresreforms},
\begin{enumerate}
\item[(a)] a total of $2^N-1$ energy storages given by 
unique functions $H_{i_1\dots i_r}$ whose compositions with $\sigma_{i_1\dots i_r}$ recover $H$
and \emph{a priori} 
undetermined energy storage trajectories ${x_{i_1\dots i_r}: \mathbb{R}\to T^*Q_{i_1}\times\dots\times T^*Q_{i_r}}$;
\item[(b)] the port spaces
$P_{i_1\dots i_r} := T^*(T^*Q_{i_1}\times \dots\times T^*Q_{i_r})\,\oplus\, T(T^*Q_{i_1}\times \dots\times T^*Q_{i_r})$
into which each energy storage can feed the dual variables $(dH_{i_1\dots i_r}){}_{x_{i_1\dots i_r}(t)}$ and $\dot x_{i_1\dots i_r}(t)$;
\item[(c)] the Dirac structure 
\begin{equation*}
  \mathscr{D}_N := \left\{  \bigolist_{\sumiall}^{\textrm{lex}} (e_{i_1\dots i_r}, f_{i_1\dots i_r})\in \bigsolder_{\sumiall}^{\textrm{lex}} \!\!P_{i_1\dots i_r}  \, \middle|\, 
  \begin{array}{ll} 
   (\mu_n^*\Omega)_\flat(f_n) = \sum\limits_{i \in \mathcal{I}}
 \left[\sigma_{i_1\dots i_r}\circ\mu_n\right]^*e_{i_1\dots i_r} 
 & \forall\, n\in\{1,\dots,N\},\\ 
   \qquad f_{i_1\dots i_r} = \sum\limits_{n=1}^N \left[\sigma_{i_1\dots i_r}\circ \mu_n\right]_*f_n &  \forall\, i \in \bigcup\limits_{r=2}^N \I{r}
  \end{array}
\right\}\,,
\end{equation*}
\end{enumerate}
 where the operation $\solder$ denotes an associative summation of bundles with overlapping base spaces which inclusively interpolates between the product bundle (no overlap) and the Whitney sum (complete overlap), which is key to our geometric approach and defined in \ref{sec:bundlemerging}, 
while the list symbol $\olist^{\textrm{lex}}_{i\in\Iall} (e_{i_1\dots i_r},f_{i_1\dots i_r})$ denotes the  lexicographically ordered tuple $(e_1,f_1, \dots, e_{12}, f_{12},\dots, e_{12\dots N}, f_{12\dots N})$. 
\newpage
\noindent The main practical result of this work is the identification of the canonical Dirac structure $\mathscr{D}_N$ and the transformation (\ref{musigmapath}) that,
starting from any canonical Hamiltonian system,
takes one there. The main structural result is our identification of the bundle sum $\solder$ that is crafted to neatly join port spaces to a total port space and to systematically reveal further essential canonical topological structure required for $\mathscr{D}_N$ which does not arise from the original Hamiltonian system.\\[6pt]
\noindent All results and constructions of this article are developed and presented for a wide readership that is only assumed to be reasonably familiar with the basics of the differential geometric description of canonical Hamiltonian systems. Readers who are additionally familiar with port-Hamiltonian theory may be able to focus directly on the admittedly technical, but essential, developments of section \ref{sec:splitmerge} and on their application and benefit to port-Hamiltonian theory in section \ref{sec:hiresreforms}. The complete organisation of this article is as follows. To set the stage for the required technical and conceptual developments, we first review, in section \ref{sec:lowres}, how one obtains a trivial port-Hamiltonian reformulation of a Hamiltonian system, which does not yet increase the resolution of detail in comparison to the Hamiltonian formulation. The transition to this trivial  reformulation  --- in marked contrast to the higher resolution reformulations we derive afterwards ---  is always possible. Although technically trivial, it reveals the two conceptual cornerstones of port-Hamiltonian theory, energy storages and Dirac structures, in a deductive rather than axiomatic way and thus presents the geometric basis from which we afterwards develop the higher resolution formulations. 
Section \ref{sec:splitmerge} then provides essential technical constructions on product manifolds with $N$ factors and develops bespoke invertible push-forwards and pull-backs along $N$-tuples of maps which provide structurally most insightful tools for the construction of splitting paths, which present the structural basis for the present work. The formulation of a bundle sum for bundles with overlapping base manifolds, which will later play a structurally essential role in the identification of the proper total port space, concludes this technical section. Section \ref{sec:hiresreforms} then 
applies the previously developed concepts and techniques for a  systematic split of the trivial energy storage and Dirac structure, which we defined for any canonical Hamiltonian system, into an exponentially large family of high-resolution energy storages and a high-resolution Dirac structure along any given factorisation of the original configuration space. The main results in this section are the appropriate construction of the total port space of the system and the identification of the maps that one ultimately recognises to effect the complete transformation from a Hamiltonian to a high-resolution port-Hamiltonian formulation.\\[6pt]
\noindent\emph{Notation:} Throughout this work, we use \emph{multi-indices} $i_1\dots i_r$ which range over the set
$$\Iall(r) := \left\{ i: \{1,...,r\} \to \{1,\dots,N\} \,\big|\,  i \textrm{ is strictly increasing}\right\} \quad\textrm{ or even }\quad \Iall := \bigcup_{r=1}^N \Iall(r)\,.$$ 
Depending on which is meant, we succinctly write either $i \in \Iall(r)$ for fixed $r$ or $i \in \Iall$ for running $r$.

\newpage
\section{Trivial port-Hamiltonian reformulation}\label{sec:lowres}
\subsection{Canonical Hamiltonian systems}\label{subsec:canham}
\noindent Throughout this article, we consider a canonical Hamiltonian system whose configuration space $M$ is given by a smooth connected finite-dimensional manifold and whose phase space is the cotangent bundle ${T^*M \stackrel{\Pi}{\longrightarrow} M}$, which carries the canonical symplectic form $\Omega\in\Omega^2(T^*M)$ that arises as the exterior derivative $\Omega=d\Pi^*$ of the pull-back of the bundle projection $\Pi$ \cite{abraham1978foundations, libermann1987symplectic}. The temporal evolution of the state of the system is encoded in a smooth curve $x: \mathbb{R} \to T^*M$ on the phase space that satisfies the Hamiltonian equation of motion
\begin{equation}\label{hamiltonianeom}
    \Omega_\flat(\dot x(t)) = dH_{x(t)}\,,
\end{equation}
where $\Omega_\flat: TT^*M \to T^*T^*M$ is the diffeomorphism defined by requiring that $\Omega_\flat(B)(A)=\Omega(A,B)$ for all $A$, $B$ in the same fibre of $TT^*M$, while the \emph{Hamiltonian} $H: T^*M \to \mathbb{R}$ is a smooth function that assigns an energy to every point of the phase space $T^*M$ and is chosen such as to model the desired trajectories for a classical mechanical system~\cite{abraham1978foundations, hamiltonianen1999introduction}. Since the Hamiltonian equation of motion only depends on $dH$, we will assume, without any loss of generality,  that $H$ contains no constant term.

\subsection{Energy storage}
\noindent The transition from this Hamiltonian formulation to the port-Hamiltonian formulation starts with the construction of an \emph{energy storage}~\cite{duindam2009, vanderschaft2014introductory}.
At the lowest level of resolution, on which we focus entirely in this section, there is only one such energy storage, given by the Hamiltonian function $H$ and an \emph{a priori} not further specified smooth trajectory $x$ on the same domain. We will find it useful to succinctly depict the constituents of this energy storage through the sequence
\begin{equation}\label{lowresstorage}
    \begin{tikzcd}
    \mathbb{R}
    \arrow[r,"x"]
    & 
    T^*\!M 
    \arrow[r,"H"]
    & 
    \mathbb{R}\,.
    \end{tikzcd}
\end{equation}
The energy storage indeed just consists of this pair of maps and neither the phase space trajectory nor the Hamiltonian will ever make an appearance in the port-Hamiltonian formulation beyond their pair-appearance in an energy storage. Within the energy storage, however, there is one construction of central conceptual relevance, namely the observation that the temporal rate of change of the energy of the system can be expressed as the natural pairing of two dual quantities, the gradient $dH_{x(t)}$ of the Hamiltonian function and the velocity $\dot x(t)$ of the trajectory, since 
\begin{equation}\label{rateofchange}
    \postdot{(H\circ x)}(t) = dH_{x(t)}(\dot x(t))
\end{equation}
due to the chain rule. Note that $dH_{x(t)} \in T^*_{x(t)}T^*M$ lives in the dual space to that of ${\dot x(t) \in T_{x(t)}T^*M}$, so that the former naturally acts on the latter, as it does on the right hand side of (\ref{rateofchange}). Since both quantities project, by the respective bundle projection, to the same base point $x(t) \in T^*M$, taken as a pair $\left((dH_{x(t)},\dot x(t)\right)$ they are an element of the Whitney sum bundle 
\begin{equation}\label{lowresportspace}
    T^*T^*M \oplus TT^*M \stackrel{\pi^\oplus}{\longrightarrow} T^*M\,,
\end{equation}
whose total space is defined as that submanifold of the product manifold $T^*T^*M \times TT^*M$ whose elements $(e,f)$ satisfy $\pi(e)=\pi(f)$, where $\pi$ denotes the respective cotangent and tangent bundle projection map, so that the Whitney sum bundle projection map $\pi^\oplus(e,f):=\pi(e)$ is well-defined. The Whitney sum bundle that arises from an energy storage is of key technical importance in port-Hamiltonian theory and constitutes the so-called \emph{port space} that arise from a storage~\cite{vanderschaft2014introductory, duindam2009}. The relevance of the port space is that it can be `fed an element' by the energy storage from which it arose. This characterisation will become clear soon. Since with (\ref{lowresstorage}), we have only one energy storage for what will turn out to be the lowest-resolution reformulation of a Hamiltonian system,  its port space (\ref{lowresportspace}) constitutes already the \emph{total port space}. Port-Hamiltonian formulations of higher resolution will contain more energy storages and, accordingly, as many port spaces that need to be tied together to one more refined total port space, but this does not concern us here yet.

\subsection{Dirac structure}
\noindent The second and only other conceptual idea underlying port-Hamiltonian theory is that of a so-called \emph{Dirac structure} \cite{courant1990dirac, vanderschaft1998affine}, which is a particular subbundle of the total port-space that shares the same base space, so that we have the diagram
\begin{equation}
\begin{tikzcd}[column sep=large, row sep=large]
    \mathscr{D}
    \arrow[r, hook, "\iota"]
    \arrow[dr, to path={ -- node[pos=0.35,left] {${\scriptstyle \left.\pi ^\oplus\right|_\mathscr{D}}$}([yshift=-8.2ex]\tikztostart.south) -- (\tikztotarget) },rounded corners=12pt]
    & 
    T^*T^*M\oplus TT^*M 
    \arrow[d,"{\pi^\oplus}" pos=0.4]
    \\
    &T^*M
\end{tikzcd}
\end{equation}
where $\iota$ denotes the canonical set inclusion map and $\mathscr{D}$ is to be specified such that the requirement
\begin{equation}\label{Diraceom}
    \left(dH_{x(t)}\,,\, \dot x(t)\right) \in \mathscr{D}
\end{equation}
is \textrm{equivalent} to the Hamiltonian equation of motion (\ref{hamiltonianeom}). Note that $\mathscr{D}$ itself does  not know anything about a Hamiltonian, or a phase space trajectory, but is just an algebraic structure constructed over the phase space of the system. Thus we must ask what the $\mathscr{D}$-defining set membership condition 
$(e,f) \in \mathscr{D}$
must be for any $(e,f)\in T^*T^*M \oplus TT^*M$ in order for (\ref{Diraceom}) to reproduce the equation of motion. It is easy to see that the choice
\begin{equation}\label{lowresD}
    \mathscr{D}:= \{(e,f)\in T^*T^*M\oplus TT^*M\,|\, \Omega_\flat(f)=e\}
\end{equation}
trivially satisfies this requirement. We emphasise that $\mathscr{D}$ just describes a particular subset of the port space. In particular, its elements do not depend on time. The port-Hamiltonian equation (\ref{Diraceom}) simply uses this fixed subset in order to constrain the relation between $dH_{x(t)}$ and $\dot x(t)$ for any time $t$ such as to reproduce the dynamics. There is a diagrammatic representation of the port-Hamiltonian equation of motion, 
\begin{equation} \label{lowresbondgraph} \includegraphics[width=0.3\linewidth, valign=c]
{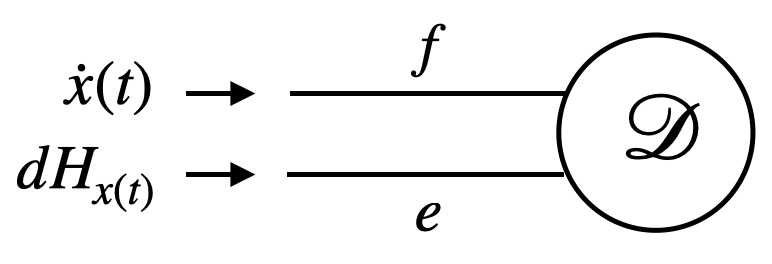}\quad,
\end{equation}
which unfolds its visual power only beyond the lowest-resolution reformulation we are considering in this section.
\noindent But its simplicity in the present case 
makes its elements particularly easy to understand by direct comparison of (\ref{lowresbondgraph}) with (\ref{Diraceom}) and (\ref{lowresD}). 

Now that we have identified the lowest-resolution Dirac structure $\mathscr{D}$ from the equation of motion for a Hamiltonian system, we note two salient properties: 
\begin{eqnarray}\label{Dprops}
    \dim \mathscr{D} = \tfrac{1}{2}\textrm{rank}(T^*T^*M \oplus TT^*M)\qquad &\textrm{ and }&\qquad
    e(f) = 0 \textrm{ for all } (e,f)\in\mathscr{D}\,. 
\end{eqnarray}
The rank property is immediate from the fact that $\dim T^*T^*M = \dim TT^*M$ and the observation that for our  $\mathscr{D}$ in (\ref{lowresD}), any $e$ is fixed by $f$. The second property also follows immediately from anti-symmetry of $\Omega$, since for $e(f) = \Omega_\flat(f)(f) = \Omega(f,f)=0$. 
While we derived these two properties only from the Dirac structure (\ref{lowresD}), which we obtained here at lowest resolution and starting from a Hamiltonian system, they indeed hold at all resolutions. To understand why this is to be expected, we note that the rank condition is necessary in order to have the set inclusion condition $(e,f)\in\mathscr{D}$ produce an equation rather than a more general or more restrictive relation and that the second condition in conjunction with (\ref{rateofchange}) and (\ref{Diraceom}) simply states that the energy of the Hamiltonian system is temporally conserved. We remark in passing  that the two properties in (\ref{Dprops}) are elevated to postulates \cite{courant1990dirac, vanderschaft2014introductory} in the general port-Hamiltonian formalism, which leads beyond the confines of closed systems, in order to ensure that a suitably defined composition of Dirac structures again results in a Dirac structure~\cite{vanderschaftintegrab, cervera2007interconnection, ortega2002interconnection}. 

A lot of mileage is gained from that, but this is well known and its exploitation is an active field of research. We will proceed in another direction: in the following section, we tailor a canonical splitting and merging formalism to underpin a subsequently performed systematic and thoroughly geometric conversion of \emph{any} Hamiltonian system
into reformulations that reveal structural detail at a resolution that grows exponentially with the number of manifolds into which the configuration space is decomposed.

\section{Splitting and merging along factorisations}\label{sec:splitmerge}
\noindent In order to arrive 
at port-Hamiltonian reformulations of higher resolution, one systematically splits the trivial port-Hamiltonian reformulation along a factorisation 
$M = Q_1 \times \dots \times Q_N$. The required splitting and merging technology for product spaces and pertinent bundles, see also \cite{yoshimura2006dirac1,lee2013introduction},
is reviewed and modestly extended in the present section. We  develop the theory for products of general manifolds $U_1\times\dots\times U_N$ in order to be able to later let $U_n$ not only be $Q_n$ but, as the need arises, also $U_n = TT^*Q_n$ for instance.

\subsection{Point splitting and merging}
 \noindent A product manifold  $U_1\times\dots \times U_N$ of $N$ smooth manifolds canonically gives rise to two families of maps. The first family are the $2^N-1$ \emph{point splitting} maps $s_{i_1\dots i_r}$ which are defined for any $i\in\Iall$
 by virtue of  
\begin{eqnarray}\label{gens}
    s^U_{i_1\dots i_r}: U_1\times\dots\times U_N &\to& U_{i_1}\times\dots\times U_{i_r}\\
    (u_1,\dots,u_N) &\mapsto& (u_{i_1},\dots,u_{i_r})\,.\nonumber
\end{eqnarray}
The second family are the $N$ different \emph{point merging maps}
defined for every $j\in\{1,\dots,N\}$ as
\begin{eqnarray}
    m^U_{j}: U_1\times\dots\times \widehat{U}_{j} \times \dots\times U_N \to (U_{j} \to U_1\times\dots\times U_N)\\
(u_1,\dots,\widehat{u}_{j},\dots,u_N) \mapsto (u_j \mapsto (u_1,\dots,u_{N}))\,,\nonumber
\end{eqnarray}
whose codomain should be noted to be a space of maps again and where a hat indicates that the respective set or element has been removed from an expression.
We will frequently encounter compositions of a point splitting map $s^U_{i_1\dots i_n}$ with a map $m^U_j(c_1,\dots,\widehat c_j,\dots, c_N)$ that arises from evaluation of the point splitting map $m_j$ at a fixed point $c_1,\dots,\widehat c_j,\dots, c_N$ of its domain, which results in the map
\begin{equation}\label{general_som}
    s^U_{i_1\dots i_r} \circ m^U_j(c_1,\dots,\widehat c_j,\dots,c_N): U_j \to U_{i_1}\times\dots\times U_{i_r}\,,\quad
    u_j \mapsto \left\{
    \begin{array}{cl}
      (c_{i_1},\dots,u_j,\dots c_{i_r})& \textrm{if } j\in\{i_1\dots i_r\}\\
      (c_{i_1},\dots\dots\dots, c_{i_r}) & \textrm{if } j\not\in \{i_1\dots i_r\}
    \end{array}
    \right\}\,,
\end{equation}
which will finally emerge in this generality, but then crucially, only in section \ref{sec:hiresreforms}. For the special case $n=1$, which will be needed throughout the paper, (\ref{general_som}) reduces to a map  $U_j \to U_i$  given by
\begin{equation}\label{som}
s^U_i\circ m^U_j(c_1,\dots,\hat c_j,\dots,c_N)=\left\{\begin{array}{cl}
  \textrm{id}_{U_i} & \textrm{if } i=j\\
  \textrm{const}_{c_i} &\textrm{if } i\neq j
  \end{array}\right\}\,,
  \end{equation}
where the constant maps $\textrm{const}_{c_i}: U_j\to U_i$ are defined as $\textrm{const}_{c_i}(u)=c_i$. Using the concise \emph{complement notation} $\overline{c_j}$ for the $(N-1)$-tuple  $(c_1,\dots,\widehat{c}_j,\dots,c_N)$, one obtains the push-forward and pull-back
\begin{equation}\label{som_pfpb}
\left(s^U_{i}\circ m^U_{j}(\overline{c_j})\right)_*= \delta_{ij}\, \textrm{id}_{TU_i} 
\quad\textrm{ and thus}\quad
\left(s^U_{i}\circ m^U_{j}(\overline{c_j})\right)^*= \delta_{ij}\, \textrm{id}_{T^*U_j}\,.
\end{equation}
In the opposite order, a composition of point merging and point splitting maps is only possible for the case $n=1$, and moreover only needed for the case $i=j$, for which one straightforwardly calculates that
$$m^U_i(\overline{c_i})\circ s^U_i= \textrm{const}_{c_1} \times \dots\times\textrm{id}_{U_i} \times\dots\times \textrm{const}_{c_N}\,,$$
so that the push-forward maps $(m^U_1(\overline{c_1})\circ s^U_1)_*\,,\dots, (m^U_N(\overline{c_N})\circ s^U_{N})_*$ are mutually orthogonal projectors on ${T_{(u_1,\dots,u_N)}(U_1\times\dots\times U_N)}$ that sum  to the identity,
\begin{equation}\label{mos_pf}
\sum_{i=1}^N(m^U_i(\overline{c_i})\circ s^U_i)_* = \textrm{id\,}_{T_{(u_1,\dots,u_N)}(U_1\times\dots\times U_N)}\,,
\end{equation}
and analogously for the pull-backs,
\begin{equation}\label{mos_pb}
\sum_{i=1}^N(m^U_i(\overline{c_i})\circ s^U_i)^* = \textrm{id\,}_{T^*_{u_1}U_1\times\dots\times T^*_{u_N}U_N}\,.
\end{equation}
 The following three subsections forge 
these point splitting and merging maps and their respective push-forwards and pull-backs into intuitively useable tools, whose judicious application will effect all required splitting and merging tasks in the following sections. 

\subsection{Splitting and merging push-forwards and pull-backs}
\noindent It is useful to concisely refer to 
$T(U_1\times\dots\times U_N)$ and $T^*(U_1\times\dots\times U_N)$ as the \emph{merged} tangent and cotangent bundles
and, analogously, to the products
$TU_1\times\dots\times TU_N$ and $T^*U_1\times\dots\times T^*U_N$ as the \emph{split} tangent and cotangent bundles. We further define the action of an element $(\alpha_1,\dots,\alpha_N)$ of the split cotangent bundle on an element $(A_1,\dots,A_N)$ of the split tangent bundle at the same base point as  ${(\alpha_1,\dots,\alpha_N)(A_1,\dots A_N):= \alpha_1(A_1)+\dots+\alpha_N(A_N)}$.

In order to be able to reversibly map the merged tangent bundle to the split tangent bundle, and vice versa, we define the \emph{splitting push-forward} $S^U{\!}_\times$ as a push-forward along the $N$-tuple of maps $(s_1,\dots,s_N)$ and the \emph{merging push-forward} $M^U{\!}_\times$ as a push-forward along the $N$-tuple of maps $(m_1,\dots,m_N)$, by virtue of letting 
\begin{equation*} 
\begin{tikzcd}[column sep=0.2cm]
      {A} \arrow[dd, maps to]
      & T_{(u_1,\dots,u_N)}(U_1\times\dots\times U_N) \arrow[dd, shift right, shift right, "{S^U{\!}_\times}",swap] 
      &       M^U{\!}_\times(A_1,\dots,A_N):=\sum\limits_{j=1}^N m^U_j(\overline{u_j})_*A_j\\
      &  & \\
      S^U{\!}_\times(A) :=\left(s^U_{1*}A,\dots,s^U_{N*}A\right)
      & T_{u_1}U_1 \times \dots \times T_{u_N}U_N \arrow[uu, shift right, shift right, "{M^U{\!}_\times}",swap]
       & {(A_1,\dots,A_N)} \arrow[uu, maps to]\,.
\end{tikzcd}
\end{equation*}
 One readily establishes that $S^U{}_\times$ and $M^U{}_\times$ are diffeomorphisms, since (\ref{som_pfpb}) and (\ref{mos_pf}) imply
\begin{eqnarray}\label{SMidTU}
     S^U{\!}_\times \circ M^U{\!}_{\times} = \textrm{id}_{TU_1\times\dots\times TU_N} 
\qquad &\textrm{and}& \quad M^U{\!}_{\times} \circ S^U{\!}_{\times} = \textrm{id}_{T(U_1\times\dots\times U_N)}\,.
\end{eqnarray}
Analogously, the \emph{merging pull-back} $M^{U\times}$ and the \emph{splitting pull-back}  $S^{U\times}$ are defined through 
\begin{equation*} 
\begin{tikzcd}[column sep=0cm]           
M^{U\times}(\alpha_1,\dots,\alpha_N) := \sum\limits_{i=1}^N s^U_i{}^{*}(\alpha_i)
      & T_{(u_1,\dots,u_N)}^*(U_1\times\dots\times U_N)  \arrow[dd, shift left, shift left, "{S^{U\times}}"]
      &  
      {\alpha} \arrow[dd, mapsto]\\
      & &  \\
      {(\alpha_1,\dots,\alpha_N)} \arrow[uu, maps to]& T_{u_1}^*U_1 \times\dots\times T_{u_N}^*U_N \arrow[uu, shift left, shift left,"M^{U\times}"] 
      & S^{U\times}(\alpha):=\left(m^U_1(\overline{u_1})^*\alpha,\dots,m^U_N(\overline{u_N})^*\alpha\right). 
\end{tikzcd}
\end{equation*} 
Also the splitting and merging pull-backs are diffeomorphisms, since due to (\ref{som_pfpb}) and now (\ref{mos_pb}),
\begin{eqnarray}\label{SMidTstarU}
S^{U\times} \circ M^{U\times} = \textrm{id}_{T^*U_1\times\dots\times T^*U_N}
\quad &\textrm{and}& \quad M^{U\times} \circ S^{U\times} = \textrm{id}_{T^*(U_1\times\dots\times U_N)}\,.
\end{eqnarray}
Note that the terminology for the maps $S^{U\times}$, $M^{U\times}$, $S^U{\!}_\times$, $M^U{\!}_\times$ and their domains and codomains is chosen such as to be able to `infer' from it that the only canonical map that maps, say,  $$T^*T^*(U_1\times\dots\times U_N) \longrightarrow T^*(T^*U_1\times\dots\times T^*U_N)$$ is given by the standard pull-back $M^{U\times*}$ of the merging pull-back $M^{U\times}$. The splitting and merging tasks one encounters in the transition from a Hamiltonian system to a port-Hamiltonian indeed become very straightforward to perform with this formalism and would be rather difficult without it.
\subsection{Splitting and merging push-forwards and pull-backs of tensors}
\noindent From the explicit definitions of $S^{U\times}$, $M^{U\times}$, $S^U{\!}_\times$, $M^U{\!}_\times$ in the previous section, one readily shows that the splitting pull-back of a merged covector $\alpha\in T_{(u_1,\dots,u_N)}^*(U_1\times\dots\times U_N)$
can be expressed in terms of the merging push-forward of a split vector $(A_1,\dots,A_N)\in T_{u_1}U_1\times\dots\times T_{u_N}U_N$ as 
\begin{equation}\label{Spb_Mpf}
(S^{U\times}\alpha)(A_1,\dots,A_N) = \alpha(M^U{\!}_{\times}(A_1,\dots,A_N))\,,
\end{equation}
while a merging pull-back of a split covector $(\alpha_1,\dots,\alpha_N) \in T_{u_1}^*U_1 \times\dots\times T_{u_N}^*U_N$ can be expressed in terms of a splitting push-forward of a merged vector $A \in T_{(u_1,\dots,u_N)}(U_1\times\dots\times U_N)$ as
\begin{equation}\label{Mpb_Spf}
(M^{\times}(\alpha_1,\dots,\alpha_N))(A) = (\alpha_1,\dots,\alpha_N)(S^U{\!}_\times A)\,.
\end{equation}
The relations (\ref{Spb_Mpf}) and (\ref{Mpb_Spf}) are not only very useful for calculations, but also already completely fix the pull-backs
$S^{U\times}$ and $M^{U\times}$ of $(0,p)$-tensors as follows:
the splitting pull-back  
$S^{U\times}$ of a 
\emph{merged $(0,p)$-tensor}
$\Omega: T_{(u_1,\dots,u_N)}(U_1\times\dots\times U_N) \times \dots \times T_{(u_1,\dots,u_N)}(U_1\times\dots\times U_N)\to \mathbb{R}$
must be defined as
\begin{equation}
    (S^{U\times} \Omega)((A_1,\dots,A_N),\dots,(Z_1,\dots,Z_N)) := \Omega(M^U{\!}_\times (A_1,\dots,A_N),\dots,M^U{\!}_\times(Z_1,\dots,Z_N))\,,
\end{equation}
In like fashion, unsurprisingly, one finds that also the merging pull-back $M^{U\times}$ of a \emph{split $(0,p)$-tensor} \allowbreak
${\Omega: (T_{u_1}U_1\times\dots\times T_{u_N}U_N) \times \dots \times (T_{u_1}U_1\times\dots\times T_{u_N}U_N)\to\mathbb{R}}$
must be defined as
\begin{equation}
    (M^{U\times} \Omega)(A,\dots, Z) := \Omega(S^U{\!}_\times A,\dots,S^U{\!}_\times Z)\,.
\end{equation}
Analogously, one may also define the action of $S^U{\!}_\times$ and $M^U{\!}_\times$ on $(p,0)$-tensors or even split and merged $(p,q)$-tensors, but none such will be required in this work. 

\subsection{Merging of bundles with overlapping base manifolds}\label{sec:bundlemerging}
\noindent Let $U_1,\dots,U_N$ be smooth manifolds. Now consider two bundles 
\begin{equation}
    \begin{tikzcd}[column sep=large, row sep=large]
        C 
        \arrow[d, "{\pi^C}"]
        & 
        D 
        \arrow[d, "{\pi^D}"]
        \\
        U_{i_1} \times\dots\times U_{i_{n_C}}  
        &
        U_{j_1} \times\dots\times U_{j_{n_D}} 
    \end{tikzcd}
\end{equation}
with smooth total spaces $C$ and $D$ and the factors of the respective base manifolds being selected by two particular sequences $i_1,\dots,i_{n_C} \in \Iall(n_C)$ and $j_1,\dots,j_{n_D}\in \Iall(n_D)$. The numbers $n_C$ and $n_D$ thus coincide with the numbers of factor manifolds in the base space of $C$ and $D$, respectively. We then define the bundle
\begin{equation}\label{solder}
\begin{tikzcd}[column sep=large, row sep=large]
        C \solder D
        \arrow[d, "{\pi^{\includegraphics[width=.48cm, height=.15cm]{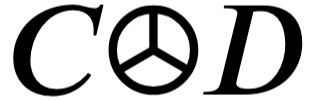}}}"]
        \\
        \quad U_{k_1}\times\dots\times U_{k_{n_{C\subscriptsolder D}}}
    \end{tikzcd}
\end{equation}
as follows. First, the total bundle space is given by \begin{equation}\label{AsolderBtotalspace}
C\solder D  := \big\{ (c,d) \in C\times D\,|\, (p_r\circ\pi^C)(c) = (p_s\circ\pi^D)(d) \textrm{ for all } (r,s)\in L_{i,j} \big\}\,,
\end{equation} 
where the locations set 
\begin{equation}\label{locationsset}
L_{i,j} := \{(r,s)\in\{1,\dots,n_C\}\times\{1,\dots,n_D\} \,|\, i_r = j_s\}\,,
\end{equation}
records, for every manifold that occurs as a common factor in both base spaces, the pair of locations of this manifold in the respective base space and the generic map $p_\ell$ yields the $\ell$-th entry of any tuple.
Second, the base space 
$$U_{k_1} \times\dots\times U_{k_{n_{C\subscriptsolder D}}}$$
is given by the strictly increasing sequence $k$ that contains every element of the set $\{i_1,\dots,i_{n_C}, j_1,\dots, j_{n_D}\}$, with $n_{C\subscriptsolder D}$ being the size of that latter set. 
Third, the bundle projection
$\pi^{C\subscriptsolder D}$ 
whose action $\pi^{C\subscriptsolder D}(c,d)$ on any $(c,d) \in C\solder D$ yields the here entry-wise defined $n_{C\subscriptsolder D}$-tuple   
\begin{equation}\label{solderedprojection}
    p_m(\pi^{C\subscriptsolder D}(c,d)):=\left\{
    \begin{array}{cl}
        (p_r\circ \pi^C)(c) & \textrm{ if } \exists r\in\{1,\dots,n_C\}: k_m = i_r  \\
        (p_s\circ\pi^D)(d) 
         & \textrm{ if } \exists s\in\{1,\dots,n_D\}: k_m = j_s
    \end{array}
    \right\}
\end{equation}
for all $m=1,\dots,n_{C\subscriptsolder D}$. Note that the projection map (\ref{solderedprojection}) is well-defined due to the conditions in (\ref{AsolderBtotalspace}) and due to the fact that the sequences $i$, $j$, $k$ are all strictly increasing. The dimension of the bundle (\ref{AsolderBtotalspace}) is clearly the dimension of the product bundle reduced by the total number of dimensions removed through the $|L_{i,j}|$ conditions, which yields 
\begin{equation}\label{solderdim}
    \dim C\solder D \,=\, \dim C \,+\, \dim D \, - \!\!\!\!\sum_{(r,s)\in L_{i,j}} \dim U_r\,.
\end{equation}
By definition of bundle rank, we have $\rank C = \dim C - \sum_{a=1}^{n_C} \dim U_{i_a}$, ${\rank D = \dim D - \sum_{b=1}^{n_D} \dim U_{j_b}}$ and ${\rank C\solder D = \dim C\solder D - \sum_{c=1}^{n_{C\subscriptsolder D}} \dim U_{k_c}}$. Substituting these into  the dimension formula (\ref{solderdim}) one finds that
\begin{equation*}
    \rank C\solder D = \rank C + \rank D + \sum_{a=1}^{n_C} \dim U_{i_a} + \sum_{b=1}^{n_D} \dim U_{j_b}
    - \!\!\!\!\sum_{(r,s)\in L_{i,j}} \dim U_r  - \sum_{c=1}^{n_{C\subscriptsolder D}} \dim U_{k_c}\,,
\end{equation*}
where the last sum cancels precisely the three preceding sums, so that one obtains the simple rank formula
\begin{equation}
    \rank C \solder D = \rank C + \rank D\,.
\end{equation}
In particular, we thus have the equalities 
\begin{equation}
    \rank C\times D = \rank C\solder D = \rank C\oplus D \,.
\end{equation}
The construction of $C\solder D$ inclusively interpolates between the limit case of a product bundle $C\times D$, to which $C\solder D$ reduces in case the base spaces of $C$ and $D$ have no common factor, and the limit case of a Whitney bundle $C\oplus D$, to which $C\solder D$ reduces in case the base spaces coincide. We remark that while the expressions for the dimensions of the cartesian product bundle $C\times D$, the Whitney bundle $C\oplus D$ and the here described bundle $C\solder D$ differ from each other, all three constructions are simply rank additive.
Between these limit cases, $C\solder D$ describes a merging of bundles that is required for the geometrically adequate definition of Dirac structures if the configuration space of a manifold is factorised into product of $N\geq 3$ manifolds. Note that, just as for the cartesian product bundle and the Whitney bundle, the operation $\solder$ is associative,
$(C \solder D) \solder E = C \solder (D \solder E)$.

\newpage
\section{High-resolution port-Hamiltonian reformulations}\label{sec:hiresreforms}
\noindent With the splitting and merging maps defined in the previous section, we are now in a position to split the lowest-order port-Hamiltonian formulation of section \ref{sec:lowres} along a factorisation $M=Q_1\times\dots\times Q_N$. More precisely, we split the low-resolution energy storage into $2^N-1$ storages and thus identify the relevant port spaces for the then-to-be-constructed split Dirac structure. Ultimately, one must split the equation of motion of the low-resolution Dirac structure and all quantities appearing in there. When the dust settles, this reveals the form of the maps $\mu_n$ and  $\sigma_{i_1\dots i_r}$ mentioned in the introduction, which for the transition from Hamiltonian to port-Hamiltonian theory play the role played of the  Legendre transformation from Hamiltonian to Lagrangian theory.

\subsection{High-resolution energy storages}
\noindent The factorisation above induces a unique split of the one low-resolution energy storage
\begin{equation}\nonumber
    \begin{tikzcd}
    \mathbb{R}
    \arrow[rr,"x"]
    & &
    T^*\!M 
    \arrow[rr,"H"]
    & &
    \mathbb{R}\,.
    \end{tikzcd}
\end{equation}
into an $\Iall$-family of $2^N-1$ high-resolution energy storages  
\begin{equation}
    \begin{tikzcd}
    \mathbb{R}
    \arrow[rr,"{x_{i_1\dots i_r}}"]
    & &
T^*Q_1\times\dots\times T^*Q_N 
    \arrow[rr,"{H_{i_1\dots i_r}}"]
    & &
    \mathbb{R}\,.
    \end{tikzcd}
\end{equation}
The energy functions $H_{i_1\dots i_r}$ and curves $x_{i_1\dots i_r}$ that constitute the high resolution storages are, indeed, uniquely determined in terms of the energy function $H$ and the curve $x$ of the low-resolution storage by virtue of the following four conditions:
\begin{enumerate}
    \item[(a)] The Hamiltonian $H: T^*(Q_1\times\dots\times Q_N) \to \mathbb{R}$ of the canonical Hamiltonian system can be split along the splitting path 
    $$
    \begin{tikzcd}
    T^*{(Q_1\times\dots\times Q_N)} 
    \arrow[r, "{S^{Q\times}}"]
    & 
    T^*Q_1 \times \dots \times T^*Q_N 
    \arrow[r, "{s^{T^*Q}_{i_1\dots i_r}}"]
    & T^*Q_{i_1} \times\dots\times T^*Q_{i_r} 
    \arrow[r,"{H_{i_1\dots i_r}}"]
    & 
    \mathbb{R}\,,
    \end{tikzcd}
    $$
    so that it can be written in terms of the energy storage functions as
\begin{equation}\label{Hdecomp}
H = \sum_{\sumiall} 
H_{i_1\dots i_r}\circ s^{T^*Q}_{i_1\dots i_r}\circ {S^{Q\times}}\,.
\end{equation}
\item[(b)] The functions of reduction rank $r$ do not contain any term that could be attributed to a function at some lower reduction rank $r'<r$. 
\item[(c)] The functions labelled by sequences of length one are all chosen to have no constant term, in accordance with our convention \footnote{See the last sentence of section \ref{subsec:canham}} to choose the Hamiltonian $H$ to contain no constant term. 
\item[(d)] The energy along a curve in the low-resolution storage equals the sum of the respective energies along the respective curves in the high-resolution energy storages,
\begin{equation}\label{totalenergy}
H\circ x = \sum_{\sumiall}  H_{i_1\dots i_r}\circ x_{i_1\dots i_r}\,.
\end{equation}
\end{enumerate}
\noindent 
From the above we note three immediate implication that will be of central importance in the identification of the Dirac structure that will connect all members of the $\Iall$-family of energy storages obtained from the above conditions. 
First, the port space arising from any individual higher-resolution energy storage can be efficiently chosen as the Whitney bundle
\begin{equation}\label{highresportspace}
P_{i_1\dots i_r} := T^*(T^*Q_{i_1}\times \dots\times T^*Q_{i_r})\,\oplus\, T(T^*Q_{i_1}\times \dots\times T^*Q_{i_r})\,,
\end{equation}
since the dual quantities
$$dH_{i_1\dots i_r} \in T^*_{x_{i_1\dots i_r}(t)}\left(T^*Q_{i_1} \times \dots \times T^*Q_{i_r}\right)
\quad\textrm{and}\quad
\dot x_{i_1\dots i_r}(t) \in T_{x_{i_1\dots i_r}(t)}\left(T^*Q_{i_1} \times \dots \times T^*Q_{i_r}\right)
$$
each project to the same base point in the shared basis manifold $T^*Q_{i_1}\times\dots\times\dots T^*Q_{i_r})$ for any $i\in\Iall$. 

Second, we note that taking the exterior derivative on both sides of (\ref{Hdecomp}) yields
\begin{equation}\label{dHdecomp}
  dH = \sum_{\sumiall} \left[s^{T^*Q}_{i_1\dots i_r} \circ S^{Q\times}\right]^*  dH_{i_1\dots i_r}
\end{equation}
and thus expresses the gradient of the Hamiltonian in terms of the gradients of the entire $\Iall$-family of energy functions $H_{i_1\dots i_r}$ of the respective high-energy storages. Note that this family of relations cannot be solved for the $dH_{i_1\dots i_r}$ in terms of $H$, although any $H_{i_1\dots i_r}$, and thus also its exterior derivative, is uniquely determined in terms of $H$ by virtue of conditions (b) and (c) above. The reason for the non-invertibility of (\ref{dHdecomp}) is that the $H_{i_1\dots i_r}$ on the right hand side implements these two conditions, while the left hand side does not. 

Third, combining conditions (\ref{Hdecomp}) and (\ref{totalenergy}) one immediately finds that the higher-resolution curves all derive directly from the low-resolution curve as
\begin{equation}\label{xii_x}
x_{i_1\dots i_r} = s^{T^*Q}_{i_1\dots i_r} \circ S^{Q\times} \circ x\,,
\end{equation}
which shows that the curves $x_{i_1\dots i_r}$ in their entirety carry precisely the same amount of information as the single curve $x$. Due to the easily ascertained injectivity of the map $s^{T^*Q}_{i_1\dots i_r}\circ S^{Q\times}$, it is clear that (\ref{xii_x}) imposes no restriction on the genericity of the curve $x_{i_1\dots i_r}$ for any particular $i\in\Iall$. Thus viewing any one particular single higher-resolution energy storage \emph{on its own}, the so induced curve $x_{i_1\dots i_r}$ is generic as long as $x$ is. 

\subsection{Construction of the total port space}
The trivial port-Hamiltonian reformulation of a canonical Hamiltonian system in section \ref{sec:lowres} featured only one energy storage, in stark contrast to the situation at hand with the $2^N-1$ high-resolution storages we identified in the previous section. In order to understand how to merge all corresponding port spaces into one \emph{total port space}, it is instructive to turn the implicit relation (\ref{xii_x}) between the curves $x_{i_1\dots i_r}$ for different $i\in \Iall$ into an explicit relation. To this end, first note from (\ref{gens}) that $s^{T^*Q}_{1\dots N}$ is just the identity on $T^*Q_1\times\dots T^*Q_N$ so that using (\ref{SMidTstarU}) one obtains $x=M^{Q\times}\circ x_{1\dots N}$ and can thus eliminate the low-resolution curve $x$ from (\ref{xii_x}), which yields  $x_{i_1\dots i_r} = s^{T^*Q}_{i_1\dots i_r}\circ x_{1\dots N}$. Since one then sees from the special cases where $r=1$ that $x_{1\dots N}=(x_1,\dots,x_N)$, one concludes that
\begin{equation}\label{factorcurve}
    x_{i_1\dots i_r} = (x_{i_1},\dots, x_{i_r})\,,
\end{equation}
which relates every curve in the $\Iall$-family $x_{i_1\dots i_r}$ to the curves $x_1,\dots, x_N$ of that family. This implies that any two high-resolution curves 
\begin{equation}
    \begin{tikzcd}[column sep=large, row sep=large]
    \mathbb{R}
    \arrow[d, "{x_{i_1\dots i_r}}"]
    & 
    & 
    \mathbb{R}
    \arrow[d, "{\bar x_{j_1\dots j_s}}"]\\
    T^*Q_{i_1}\times\dots\times T^*Q_{i_r}
    &
    &
    T^*Q_{j_1}\times\dots\times T^*Q_{j_s}
    \end{tikzcd}
\end{equation}
prescribe precisely the same curve on all factor manifolds on which they happen to overlap. More precisely, using that due to (\ref{factorcurve}) one has
$x_{i_1\dots i_r} = (x_{i_r},\dots,x_{i_r})$ and $\bar x_{j_1\dots j_s} = (\bar x_{j_1},\dots,\bar x_{j_s})$, 
it follows that 
$$x_{i_{a}} = \bar x_{j_b} \qquad \textrm{for all } (a,b) \in L_{i,j}\,,$$
where $L_{i,j}$ denotes the locations set defined in (\ref{locationsset}). 

It should be noted that the velocity curve $\dot x_{i_1\dots i_r}$ of an energy storage trajectory \emph{cannot be} given directly by $(\dot x_{i_1},\dots,\dot x_{i_r})$, since the former is a curve in $T(T^*Q_1\times\dots\times T^*Q_N)$ and the latter is a curve on ${TT^*Q_1\times\dots\times TT^*Q_N}$. To straightforwardly obtain the correct relation, one starts from the time derivative
\begin{equation}\label{dxii_x}
\dot x_{i_1\dots i_r} = (s^{T^*Q}_{i_1\dots i_r} \,\circ \,S^{Q\times})_* \,\circ \, \dot x
\end{equation}
of (\ref{xii_x}) again 
and considers the special cases $\dot x_i = (s_i^{T^*Q} \,\circ\, S^{Q\times})_* \,\circ\, \dot x$ for $r=1$ and $i=1,\dots, N$, which can be summarised as
\begin{equation}\label{dotxii_x}
    (\dot x_1, \dots, \dot x_N) = S^{T^*Q}{}_\times \circ S^{Q\times}{}_* \circ \dot x
\end{equation}
due to the definition of the splitting push-forward $S^{T^*Q}{}_\times$. But then with (\ref{SMidTU}) and (\ref{SMidTstarU}) one can solve the previous relation for
$\dot x$ 
and insert the result back into the general relation (\ref{dxii_x}) to obtain 
\begin{equation}\label{dxrels}
    \dot x_{i_1\dots i_r} = \big(s^{T^*Q}_{i_1\dots i_r}{}_* \circ M^{T^*Q}{}_\times\big)\circ (\dot x_1,\dots,\dot x_N) 
\end{equation}
which now encapsulates the  relations among all velocity curves $\dot x_{i_1\dots i_r}$ without any reference anymore to the velocity $\dot x$ of the low-resolution curve $x$. In conjunction with (\ref{factorcurve}) it is trivial to note that the tangent bundle projection $\pi$ from  $T(T^*Q_{i_1}\times \dots\times T^*Q_{i_r})$ takes the velocity curve to the curve
$\pi \circ \dot x_{i_1\dots i_r} = (x_{i_1},\dots,x_{i_r})$.
It thus follows for any  $i,j\in\Iall$ that the pair of velocity curves $(\dot x_{i_1\dots i_r},\dot x_{j_1\dots j_s})$ lies in the total space of the bundle
\begin{equation}
\begin{tikzcd}[row sep = large]
     T(T^*Q_{i_1}\times\dots\times T^*Q_{i_r}) \,\solder\,  T(T^*Q_{j_1}\times\dots\times T^*Q_{j_s})
     \arrow[d,"{\pi^{\includegraphics[width=.18cm, height=.18cm]{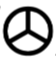}}}" pos = 0.4]
     \\
     T^*Q_{k_1}\times\dots\times T^*Q_{k_n}
\end{tikzcd}
\end{equation}
where $k$ is a strictly increasing sequence that contains every element of the set $\{i_1,\dots,i_r, j_1, \dots, j_s\}$ with $n$ being the size of that latter set; see section \ref{sec:bundlemerging}. But then also for any $i,j\in\Iall$
$$\left((dH_{i_1\dots i_r})_{x_{i_1\dots i_r}(t)}\,,\, \dot x_{i_1\dots i_r}(t)\,,\, (dH_{j_1\dots j_s})_{x_{j_1\dots j_s}(t)}\,,\, \dot x_{j_1\dots i_s}(t)\right)$$
lies in the total space of the bundle
\begin{equation}
    \begin{tikzcd}[row sep=large]
    P_{i_1\dots i_r} \,\solder\, P_{j_1\dots j_s}
     \arrow[d,"{\pi^{\includegraphics[width=.18cm, height=.18cm]{subscriptsolderpic.png}}}" pos = 0.4]
     \\
     T^*Q_{k_1}\times\dots\times T^*Q_{k_n}
     \end{tikzcd}
\end{equation}
with the same base space. Thus follow two key technical observations of this paper: First, port spaces of different energy storages \emph{can} be merged without unnecessarily inflating the base space in contrast to the product bundle. Choosing to do so then automatically implements the relation (\ref{factorcurve}) explicitly and the relation (\ref{dxrels}) implicitly. Thus if one constructs the total port space in this fashion, one recognises that for any element
\begin{equation}\label{totalps}
    \bigolist_{\sumiall}^{\textrm{lex}} (e_{i_1\dots i_r}, f_{i_1\dots i_r}) \quad\in\quad \bigsolder_{\sumiall}^{\textrm{lex}} P_{i_1\dots i_r}
\end{equation}
that could possibly be an element of a Dirac structure connecting these energy storages, one must have that
\begin{equation}\label{dxrelsf}
    f_{i_1\dots i_r} = \big(s^{T^*Q}_{i_1\dots i_r}{}_* \circ M^{T^*Q}{}_\times\big)\circ ( f_1,\dots,f_N)\qquad \textrm{for all } i \in \bigcup_{r=2}^N \I{r}\,,
\end{equation}
which simply enforces (\ref{dxrels}) in accordance with (\ref{factorcurve}). Note that we excluded the tautological cases  ${r=1}$.

\newpage
\subsection{High-resolution Dirac structure}
\noindent 
\noindent We are finally ready to show  how the factorisation $M=Q_1\times \dots\times Q_N$ also induces a unique splitting of the Dirac structure $\mathscr{D}$ of low-resolution 
\begin{center}
\begin{tikzpicture}
    \draw (0, 0) node[inner sep=0] {\includegraphics[width=.65\linewidth]{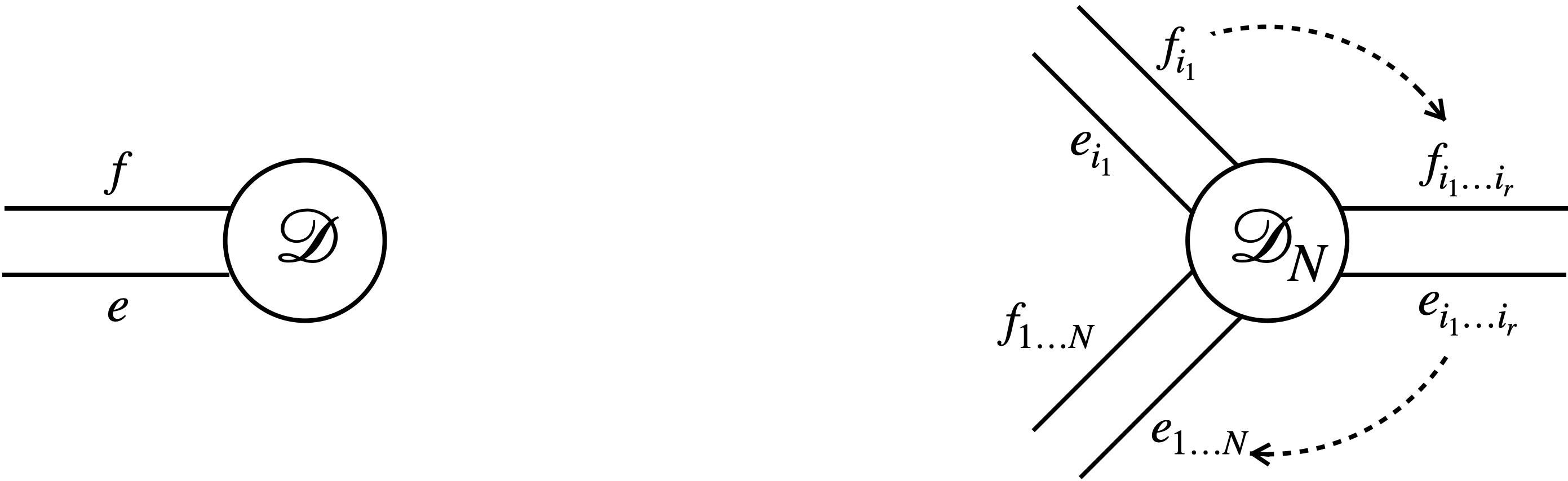}};
    \draw (-0.7, 0) node {into a Dirac structure};
\end{tikzpicture}
\end{center}
\vspace{-12pt}
of high resolution, which turns out to be uniquely determined by virtue of the following three conditions:
\begin{enumerate}
    \item The total port space is the merged bundle
    \begin{equation}
        \bigsolder_{\sumiall}^{\textrm{lex}} P_{i_1\dots i_r}
    \end{equation}
    of the $\mathcal{I}$-family of port spaces $P_{i_1\dots i_r}$ with overlapping base spaces.
    \item The Dirac structure $\mathscr{D}_N$ is the subbundle 
    \vspace{-12pt}
$$\begin{tikzcd}[column sep=large, row sep=large]
    \mathscr{D}_N
    \arrow[r, hook]
    \arrow[dr, to path={ -- node[pos=0.46,left] {${\scriptstyle \left.\pi^{\includegraphics[width=.18cm, height=.18cm]{subscriptsolderpic.png}\includegraphics[width=.18cm, height=.18cm]{subscriptsolderpic.png}}\right|_\mathscr{D}}$}([yshift=-10.1ex]\tikztostart.south) -- (\tikztotarget) },rounded corners=12pt]
    & 
    \bigsolder_{\sumiall}^{\textrm{lex}} P_{i_1\dots i_r} 
    \arrow[d, swap, shift right, shift right, shift right, "{\pi^{\includegraphics[width=.18cm, height=.18cm]{subscriptsolderpic.png}\includegraphics[width=.18cm, height=.18cm]{subscriptsolderpic.png}}}" pos=0.32]
    \\
    &T^*Q_1\times\dots\times T^*Q_N \quad.
\end{tikzcd}$$
   determined by the requirement that 
\begin{equation}\label{lowhighequiv}
    \centering
\includegraphics[width=0.90\linewidth, valign=c]{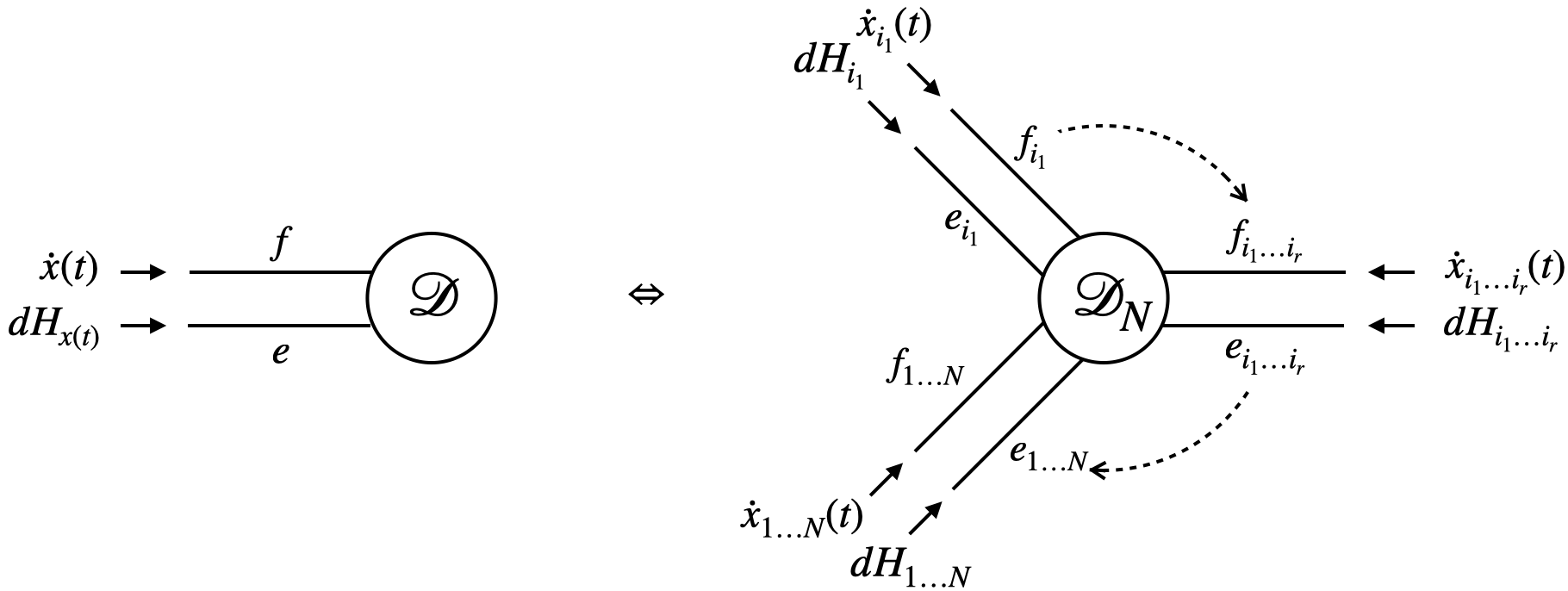}
\end{equation}
\end{enumerate}
\noindent The first conditions require no further work, but simply execution. In order to obtain $\mathscr{D}_N$ such as to satisfy the second condition, we split the equations of motion (\ref{hamiltonianeom}), which are represented here by the diagram on the left hand side of (\ref{lowhighequiv}) -- together with all quantities $\Omega$, $\dot x(t)$ and $dH_{x(t)}$ appearing therein -- along the splitting path
\begin{equation}\label{splitpath}
\begin{tikzcd}[column sep=3.5em, /tikz/every node/.append style={outer xsep=4.5pt}]
    T^*T^*{(Q_1\times\dots\times Q_N)} 
    \arrow[r, "{M^{Q\times}*}", start anchor=east, end anchor=west, shorten <=-5pt, shorten >=-5pt]
    & 
    T^*(T^*Q_1 \times \dots \times T^*Q_N )
    \arrow[r, "{S^{T^*Q \times }}", start anchor=east, end anchor=west, shorten <=-5pt, shorten >=-5pt]
    & T^*T^*Q_{1} \times\dots\times T^*T^*Q_{N}\,.
    \end{tikzcd}
\end{equation}
\noindent First, for the canonical symplectic structure $\Omega=d\Pi^*$ on $T^*M$, the above splitting yields
\begin{equation}\label{bigandsmallomega}
    (S^{T^*Q\times}\circ M^{Q\times*})\,\Omega = (\omega_1,\dots,\omega_N)\,,
\end{equation}
where $\omega_1,\dots,\omega_n$ are the canonical symplectic structures on $T^*Q_1,\dots,T^*Q_N$. To see this, it obviously suffices to  
  determine how the canonical one-form $\Pi^*$ on $T^*(Q_1\times\dots\times Q_N)$ arises from a merging of the canonical one-forms $\pi_1^*,\dots,\pi_N^*$ on the respective factor phase spaces $T^*Q_1, \dots, T^*Q_N$. It is well-known that this is always possible and that this merge reads, in our language,
\begin{equation}\label{canoneform}
  \Pi^* = (S^{Q\times*}\circ M^{T^*Q\times})(\pi_1^*,\dots,\pi^*_N)\,.
  \end{equation}
  Indeed, constructing the splitting path between the quantities that appear in the outer left column and the outer right column in the below diagram, using the by now familiar strategy to insert any required intermediate spaces in the middle column and to then make the necessary choices of splitting and merging maps along those paths, yields 
\begin{equation}\label{bigdiagram}
    \begin{tikzcd}
        T_P^*T_{(q_1,\dots,q_N)}^*(Q_1\times\dots\times Q_N) 
        & &
        T_{(p_1,\dots,p_N)}^*(T_{q_1}^*Q_1\times\dots\times T_{q_N}^*Q_N)
        \arrow[ll,swap,"{S^{Q\times*}}"]
        & &
        T_{p_n}^*T_{q_n}^*Q_n 
        \arrow[ll,swap,"{s_n^{T^*Q*}}"]
        \\
        & &  
        & &
        \\
        T_{(q_1,\dots,q_N)}^*(Q_1\times\dots\times Q_N)
          \arrow[uu,"{\Pi^*}"]
        \arrow[rr,"{S^{Q\times}}"]
        \arrow[ddrr, to path={ -- node[pos=0.4,left] {${\scriptstyle\Pi}$}([yshift=-10ex]\tikztostart.south) -- (\tikztotarget) },rounded corners=12pt, "\Pi"]
        & & 
        T_{q_1}^*Q_1 \times\dots\times T_{q_N}^*Q_N
        \arrow[uu,swap,"{M^{T^*Q\times}(\pi_1^*,\dots,\pi_N^*)}"]
        \arrow[rr,"{s_n^{T^*Q}}"]
        \arrow[dd,"{\pi\times\dots\times\pi}"]
        & &
        T_{q_n}^*Q
        \arrow[uu,swap,"{\pi^*}"]
        \arrow[dd,"{\pi}"]
        \\
        & &  
        & &
        \\
        & &  
        Q_1\times\dots\times Q_N 
        \arrow[rr,shift left,"{s^Q_n}"]
        & &
        Q
    \end{tikzcd}
\end{equation}
where $n\in\{1,\dots,N\}$ in the last column and $(p_1,\dots,p_N):=S^{Q\times}(P)$. We note, once again, that with the conceptually required start points and end points of the splitting paths in this diagram fixed, the choice of splitting maps connecting them is unique. It is then just a straightforward exercise to show that the diagram commutes.
We again emphasise that this result is known from the literature \cite{abraham1978foundations,yoshimura2006dirac1} and include it here in terms of our splitting and merging notation in order for the reader to be able to seamlessly connect it to the calculations in this paper.

Second, we split the Hamiltonian equation of motion (\ref{hamiltonianeom}) itself along the splitting path (\ref{splitpath}) at any given time $t\in\mathbb{R}$. In order to analyse the left hand side of the split equation,
\begin{equation}\label{woAA}
    (S^{T^*Q \times} \circ M^{Q \times *})(\Omega_\flat(\dot x(t)))\,, 
\end{equation}
we apply it to an arbitrary split vector which we will remove again later for comparison to (\ref{woAA}). This yields
$$
\Omega\left(M^{Q\times}{}_* (M^{T^*Q}{}_\times (A_1, \dots, A_N))\,, f
\right)  \qquad \textrm{for any }(A_1, \dots, A_N) \in TT^*Q_1 \times \dots \times  TT^*Q_N 
$$
after having used (\ref{Spb_Mpf}) to express the splitting pull-back in terms of the merging push-forward and then the definition of a standard pull-back and finally the definition of $\Omega_\flat$ in terms of $\Omega$.
Then by substituting $\Omega$ using (\ref{bigandsmallomega}) together with (\ref{SMidTstarU}), expressing the pull-backs in terms of push-forwards and using the identity (\ref{SMidTU}), one further  obtains
$$
\sum_{n=1}^N \omega_n \left( 
s_{n}^{T^*Q}{}_{*} \circ M^{T^*Q}{}_\times(A_1, \dots, A_N), 
(s_{n}^{T^*Q} \circ S^{Q\times}{})_*\dot x(t) 
\right)\,.
$$
Using first the definition of a merging push-forward and the identity (\ref{som})
in the first slot, then relation (\ref{dotxii_x}) together with the explicit definition of $S^{T^*Q}{}_\times$ in the second slot, and finally 
a definition of $\omega_{n\flat}$ in terms of $\omega_n$ analogous to that of $\Omega_\flat$ in terms of $\Omega$, yields
$$
\sum_{n=1}^N \left(\omega_{n\flat}    (\dot x_n(t) )
\right)(A_n) \,\, =\,\, \left(\omega_{1 \flat}(\dot x_1(t)), \dots, \omega_{N\flat}(\dot x_N(t)) \right)(A_1,\dots,A_N)\,,  
$$
where the equality holds due to our definition of a split element of cotangent bundle on an element of the split tangent bundle. The announced removal of the arbitrary split vector, on which we chose to act on in the beginning, thus yields the final form
\begin{equation}\label{eom:lhs}
\left(\omega_{1 \flat}(\dot x_1(t)), \dots, \omega_{N\flat}(\dot x_N(t)) \right) 
\end{equation}
for the split left-hand side of equations of motion. To split the right-hand side of (\ref{hamiltonianeom}) accordingly, consider
\begin{equation}
   (S^{T^*Q\times} \circ M^{Q \times*})\,dH_{x(t)}
\end{equation}
which, using relation (\ref{Hdecomp}),
amounts to
$$
(S^{T^*Q\times} \circ M^{Q \times*})\,e=
\sum_{\sumiall} (S^{T^*Q \times } \circ s^{T^*Q}_{i_1 \dots i_r}{}^*)\,(dH_{i_1 \dots i_r})_{x_{i_1\dots i_r}(t)} \;,
$$
and thus, by using the explicit definition of the splitting pull-back, further to
\begin{equation}\label{eom:rhs}
    \sum_{\sumiall} \left(
(s^{T^*Q}_{i_1 \dots i_r} \circ m_1^{T^*Q}(\overline{\pi(e_1)})^*(dH_{i_1 \dots i_r})_{x_{i_1\dots i_r}(t)}, \dots, (s^{T^*Q}_{i_1 \dots i_r} \circ m_N^{T^*Q}(\overline{\pi(e_N)})^*(dH_{i_1 \dots i_r})_{x_{i_1\dots i_r}(t)}
\right) \,.
\end{equation}
Comparing (\ref{eom:lhs}) and (\ref{eom:rhs}) componentwise, one obtains the split Hamiltonian equations of motion in form of $N$ coupled equations  
\begin{equation}
\omega_{n \flat}(f_n) = \sum_{\sumiall} \left( 
s_{i_1 \dots i_r}^{T^*Q} \circ m_{n}^{T^*Q}(\overline{\pi(e_n))}
\right)^* \,e_{i_1\dots i_r} \quad \textrm{for } n=1, \dots, N\,.
\end{equation}
The requirement that conditions (\ref{lowhighequiv}) and (\ref{dxrels}) are both fulfilled thus leads to the unique high-resolution Dirac structure
\begin{equation}\label{hiresDleaner}
  \mathscr{D}_N := \left\{ \bigolist_{\sumiall}^{\textrm{lex}} (e_{i_1\dots i_r}, f_{i_1\dots i_r}) \in \bigsolder_{\sumiall}^{\textrm{lex}} P_{i_1\dots i_r}\, \middle|\, 
  \begin{array}{ll} 
  {\omega_n}_\flat(f_n) =   \sum\limits_{\sumiall}
\left[s^{T^*Q}_{i_1\dots i_r}\circ
m_n^{T^*Q}(\overline{\pi(e_n)})\right]^* \,e_{i_1\dots i_r}
  & \forall\, n \in\{1,\dots,N\}
  \\
  \,\,\, f_{i_1\dots i_r} = \,\,\sum\limits_{n=1}^N \left[s^{T^*Q}_{i_1\dots i_r} \circ m^{T^*Q}_n(\overline{\pi(f_n)})\right]_* f_n\quad& \forall\, i\,\in\,\bigcup\limits_{r=2}^N \I{r}
  \end{array}\right\}
\end{equation}
for the canonical Hamiltonian system (\ref{hamiltonianeom}). It is straightforward to ascertain that the high-resolution Dirac structure $\mathscr{D}_N$ still features the two salient properties 
\begin{equation}
    \rank \mathscr{D}_N = \tfrac{1}{2} \rank P_N \qquad\textrm{ and }\qquad \sum_{\sumiall}  e_{i_1\dots i_r}(f_{i_1\dots i_r}) = 0 \quad \textrm{ for all } \bigolist_{\sumiall}^{\textrm{lex}} (e_{i_1\dots i_r}, f_{i_1\dots i_r}) \in \mathscr{D}_N\,.
\end{equation}
Indeed, using the invertibility of the maps ${\omega_n}_\flat$, inspection of the conditions in the above definition of $\mathscr{D}_N$ reveals that all $f_{i_1\dots i_r}$ are determined in terms of all $e_{i_1\dots i_r}$, so the rank condition is manifestly satisfied. The second property is confirmed by direct calculation.

The above obtained form (\ref{hiresDleaner}) of the high-resolution Dirac structure $\mathscr{D}_N$ allows us to finally read off the definition of the families of maps $\mu_n$ and $\sigma_{i_1\dots i_r}$, which we advertised in the introduction as the analogue of the Legendre transformation for the transition from Hamiltonian to Lagrangian mechanics, only that these functions effect the transition to port-Hamiltonian mechanics. 
They are given by the commutative diagram 
\begin{equation}\label{musigmadef}
    \begin{tikzcd}[row sep=large, column sep=small]
    T^*Q_n 
    \arrow[rr,"{\mu_n}"]
    \arrow[dr,"{m_n^{T^*Q}}"]
    &  
    &
    T^*(Q_1\times\dots\times Q_N)
    \arrow[rr, "{\sigma_{i_1\dots i_r}}"]
    \arrow[dr,"{S^{Q\times}}"]
    & 
    &
    T^*Q_{i_1} \times\dots\times T^*Q_{i_r}\\
    &
    T^*Q_1\times\dots\times T^*Q_N
    \arrow[ur,"{M^{Q\times}}"]
    &
    &
    T^*Q_1\times\dots\times T^*Q_N
    \arrow[ur, "{s^{T^*Q}_{i_1\dots i_r}}"]
    &
    \end{tikzcd}
\end{equation}
which reveals that their composite use $\sigma_{i_1\dots i_r}\circ \mu_n$ in (\ref{hiresDleaner}) amounts to a \emph{four-step} decomposition path. 

One may appreciate that 
the fully geometric definition of the high-resolution Dirac structure $\mathscr{D}_N$ would have been hard to guess without systematic splitting and merging. This is particularly the case for those defining relations that relate all $f_{i_1\dots i_r}$ to each other, since these do not follow from the original canonical Hamiltonian equations but from the careful identification of the total port space (\ref{totalps}) in terms of a $\solder$ sum.

It is interesting to note that the high-resolution Dirac structure is governed entirely by the configuration manifold's topology, with the exception of the use of the canonical symplectic forms, which also depend on the smooth and canonical structure of the cotangent bundles of the factor manifolds $Q_1,\dots, Q_N$. Given the decomposition (\ref{introfac}), $\mathscr{D}_N$ is thus entirely canonical.

\subsection{Isolation of the pure interconnection Dirac structure}\label{sec:refinement}
\noindent Having obtained the high-resolution Dirac structure $\mathscr{D}_N$ we now further decompose it in a way that separates the purely topologically determined part of the Dirac structure from the parts given by canonical symplectic forms. The strategy is to represent $\mathscr{D}_N$ as a composition of the $2^N$ Dirac structures  shown in the last diagram of this subsection, 
delegating the canonical symplectic structures $\omega_1,\dots, \omega_N$ into the eponymous Dirac structures, and to then identify $\mathscr{D}_{int}$ as the pure interconnection structure at the centre. 

The technology to do this comes from port-Hamiltonian theory, where a system is viewed as a network of interconnected \emph{open} port-Hamiltonian systems~\cite{duindam2009, vanderschaft2014introductory}. Building on the definition from Sec. \ref{sec:lowres}, where a port-Hamiltonian system consists of an energy storage on $T^*M$ and a Dirac structure, the description of an open system is characterised by a Dirac structure $\mathscr{D}$ on an extended subbundle of
$$T^*T^*M \oplus TT^*M \oplus E^*\oplus E \stackrel{\pi^\oplus}{\longrightarrow} T^*M$$ encompassing not only the port space of energy storage variables, but also \emph{external port space} defined as a Whitney sum $E^*\oplus E$ of a vector bundle
 $E \stackrel{\Pi}{\longrightarrow} T^*M$ and its dual bundle $E^* \stackrel{\Pi}{\longrightarrow} T^*M$. Though external port space may coincide with the energy storage port space, its variables $(u,y) \in E^*\oplus E$ serve a distinct role: they model interactions between a system and its environment~\cite{duindam2009, maschke1992port}. 
Given the Hamiltonian $H$, equations of motion of an open port-Hamiltonian system are then specified by requiring 
$$\big( 
dH_{x(t)}, \dot x(t), u(t), y(t)
\big) \in \mathscr{D} \,, $$ 
which may be represented diagrammatically as
\begin{equation} 
\label{openbondgraph} 
\includegraphics[width=0.38\linewidth, valign=c]
{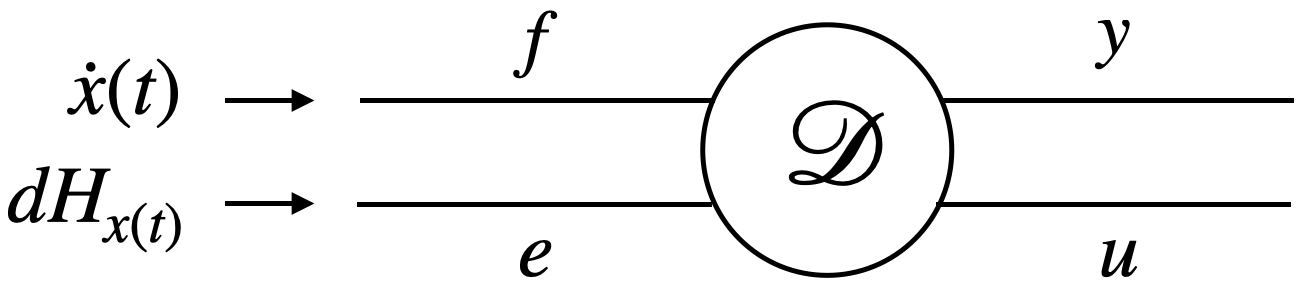}\quad.
\end{equation}
Crucially, curves $\dot x$, $u$ and $y$ are not known \emph{a priori}. Since $\mathscr{D}$ satisfies the rank condition (\ref{Dprops}), specifying the Hamiltonian is insufficient to yield a fully determined system of equations. 
Determining the evolution of $x$ therefore requires additional relations on the external port variables $(u,y)$, hence the designation of the system as \emph{open}. While such relations may be externally enforced, the advantage of the port-Hamiltonian formulation lies in supplying them through interconnections between open systems. Furthermore, the second defining property (\ref{Dprops}) of the Dirac structure dictates the power balance equation along trajectory $x$,
$$
\postdot{(H \circ x)}(t) + u(t) \left( y(t) \right)=0 \,,
$$
where we used (\ref{rateofchange}) to equate the pairing between energy storage variables to the rate of change of energy. Rather than vanishing, this rate of change is balanced by the external power flow given by the pairing of external ports. 

In the present setting, we regard each of the $N$ canonical Hamiltonian systems encoded in the high-resolution Dirac structure $\mathscr{D}_N$	of (\ref{hiresDleaner}) as an open system, as an intermediate step in refining the closed port-Hamiltonian system.
Specifically, we rewrite the defining relations of $\mathscr{D}_N$ by introducing auxiliary port variables, thereby separating the equations of motion governing the individual canonical Hamiltonian systems from those describing their mutual interaction. This yields a family of constituent Dirac structures, one for each canonical Hamiltonian system and each higher-rank energy storage tied together by an interconnecting Dirac structure that encodes their coupling. Together, this interconnected network of Dirac structures fully recovers the original high-resolution equations of motion.

We first look at the first $N$ relations in (\ref{hiresDleaner}), describing the high-resolution equations of motion. We separate off the $r=1$ term which encodes the stand-alone contribution of the $n$-th system, from the higher-rank interaction term, $r \leq 2$,
\begin{equation} \label{uncoupled}
{\omega_n}_\flat(f_n) =  e_n + u_n \,, 
\end{equation}
where we used (\ref{som}) to simplify the composition on the right-hand side of (\ref{hiresDleaner}) for $r=1$
and replaced the remaining sum over $r\geq 2$ terms by $u_n \in T^*T^*Q_n$. 
To cast the decoupled equation (\ref{uncoupled}) into a port-Hamiltonian form, we must identify the extended port space and its underlying Dirac structure. 
While the variables $(e_n,f_n)$ appearing in (\ref{uncoupled}) already constitute a dual pair, accommodating the additional variable $u_n$ requires introducing 
 a corresponding dual variable $y_n \in TT^*Q_n$.
We now construct a subbundle of $P_n \oplus P_n$ consisting of all quadruples $(e_n, f_n, u_n, y_n)$ that satisfy (\ref{uncoupled}). 
The variable $y_n$ is determined by requiring that the second defining property of a Dirac structure (\ref{Dprops}),
$$
e_n(f_n) + u_n(y_n)=0 \,,
$$
holds for all elements of the subbundle.
Solving (\ref{uncoupled}) for $e_n$, substituting it into this equation, exploiting the anti-symmetry of $\omega_{n\flat}$ and linearity reduces the equation to 
$$
u_n(-f_n + y_n) = 0 \,.
$$
Since this identity must hold for every $u_n$, it follows that $y_n=f_n$.
This equation, together with (\ref{uncoupled}) allows us to define $N$ Dirac structures, one for each canonical Hamiltonian system,
\begin{equation}\label{littleomD}
  \mathscr{D}_{\omega_n} := \left\{ \;\vphantom{\rule{0pt}{22pt}} (e_{n}, f_{n}, u_n, y_n) \in P_n \oplus P_n \; \middle|\; 
  \begin{aligned}[c]
  {\omega_n}_\flat(f_n) &= e_n+ u_n \,, 
  \\
  y_n &= f_n
  \end{aligned} \;\right\} \quad \textrm{for } n=1, \dots, N \,.
\end{equation}
The rank property (\ref{Dprops}) of the Dirac structure is satisfied by observing that any choice of $e_n$ and $u_n$ fixes the remaining variables $f_n$ and $y_n$, so the fibre is parametrised by exactly half the total variables of the port space $P_n \oplus P_n$, whereas the second property is satisfied by construction. We again emphasize that requirement
$$
\big((dH_{n})_{x_{n}(t)}, \dot x_n(t), u_n(t), y_n(t) \big) \in \mathscr{D}_{\omega_n}
$$
does not, on its own, recover the original high-resolution equations of motion of the $n$-th canonical Hamiltonian system. Indeed, $u_n$ is deliberately left unspecified at this stage so that $\mathscr{D}_{\omega_n}$ describes an open system. Were $u_n$ identified with the sum over higher-rank interaction terms present in (\ref{hiresDleaner}), the equations of motion would be restored. However, rather than imposing these missing relations externally, they are collected separately into an interconnecting Dirac structure given below, which, when joined with the constituent Dirac structures recovers the original high-resolution Dirac structure and hence the equations of motion.

The same construction applies to the second relation in (\ref{hiresDleaner}), which describes relations among trajectories of energy storages of rank $r \geq 2$. We replace the summation on the right-hand side by auxiliary variable $y_{i_1\dots i_r} \in T(T^*Q_{i_1} \times \dots \times T^*Q_{i_r})$ and thus obtain a simple identity
\begin{equation} \label{refinedfis} 
f_{i_1\dots i_r} = y_{i_1\dots i_r} \,.
\end{equation}
Introducing the variables $e_{i_1 \dots i_r}$ and $u_{i_1 \dots i_r}$ that are dual to $f_{i_1 \dots i_r}$ and $y_{i_1 \dots i_r}$, respectively, we define a Dirac structure of the bundle $P_{i_1\dots i_r} \oplus P_{i_1\dots i_r}$  by requiring that the second defining property of (\ref{Dprops}) holds. Using (\ref{refinedfis}), we obtain
$(e_{i_1\dots i_r} + u_{i_1 \dots i_r})(f_{i_1 \dots i_r}) = 0 \,,$
which holds for every $f_{i_1 \dots i_r}$. This implies $u_{i_1\dots i_r} =- e_{i_1 \dots i_r}$ and yields, together with (\ref{refinedfis}),  a Dirac structure for any strictly increasing sequence of indices $i_1, \dots, i_r$ where $r \geq 2$,
\begin{equation}\label{onejuncD}
  \mathscr{D}_{i_1\dots i_r} := \left\{ \; \vphantom{\rule{0pt}{22pt}} (e_{i_1\dots i_r}, f_{i_1\dots i_r},u_{i_1\dots i_r}, y_{i_1\dots i_r}) \in P_{i_1\dots i_r}\oplus P_{i_1\dots i_r} \; \middle|\; 
  \begin{aligned}[c]
    f_{i_1 \dots i_r} &= y_{i_1\dots i_r} \\
   u_{i_1\dots i_r} &= -e_{i_1\dots i_r} \,, 
  \end{aligned} \; \right\} \quad \textrm{for } r=2, \dots, N \,.
\end{equation}
As in the preceding construction, the interaction with other systems is encoded by the external ports $(u_{i_1 \dots i_r}, y_{i_1 \dots i_r})$.
The remaining relations between the variables are collected into a single  Dirac structure 
\begin{equation}\label{interD}
  \mathscr{D}_{int} := \left\{ \bigolist_{\sumiall}^{\textrm{lex}} (v_{i_1\dots i_r}, z_{i_1\dots i_r}) \in P \, \middle|\, 
  \begin{array}{ll} 
  \,\,\,\, - v_n =  \sum\limits_{\sumiall}
\left[s^{T^*Q}_{i_1\dots i_r}\circ
m_n^{T^*Q}(\overline{\pi(v_n)})\right]^* \,v_{i_1\dots i_r} & \forall n\in\{1,\dots,N\}
  \\
   z_{i_1\dots i_r} =\,\, \sum\limits_{n=1}^N \left[s^{T^*Q}_{i_1\dots i_r} \circ m^{T^*Q}_n(\overline{\pi(y_n)})\right]_* z_n & \forall\,i\in\bigcup\limits_{r=2}^N \I{r}
  \end{array}\right\} \,,
\end{equation}
where $P$ denotes the total port space (\ref{totalps}). 
The role of $\mathscr{D}_{int}$ is to interconnect the open constituent Dirac structures  $\mathscr{D}_{\omega_n}$ and $\mathscr{D}_{i_1 \dots i_r}$ by imposing precisely those relations that were formerly encoded in the coupling terms of the high-resolution Dirac structure. Note that while  the second defining property (\ref{Dprops}) requires a minus sign between dual port relations, its specific placement in the equation for $v_n$, rather than $z_{i_1 \dots i_r}$ is dictated by the requirement to recover the original coupling relations. The refined description of 
 $\mathscr{D}_N$ is therefore obtained by requiring the auxiliary port variables
$\bigolist_{\sumiall}^{\textrm{lex}} (u_{i_1\dots i_r}, y_{i_1\dots i_r})$
to satisfy all of the relations imposed by $\mathscr{D}_{\omega_n}$,  $\mathscr{D}_{i_1\dots i_r}$ and the interconnecting Dirac structure $\mathscr{D}_{int}$, simultaneously,
\begin{equation}\label{totD}
  \left\{ \bigolist_{\sumiall}^{\textrm{lex}}(e_{i_1\dots i_r}, f_{i_1\dots i_r}) \in P_N \, \middle|\, 
  \begin{array}{ll} 
    \raisebox{-.4ex}{\scalebox{1.3}{$\exists$}}\; \bigolist_{\sumiall}^{\textrm{lex}} (u_{i_1\dots i_r}, y_{i_1\dots i_r}) \in P_N   \,\, \textrm{ such that} &
    \\
    \,\,\,\Big[\bigolist_{\sumiall}^{\textrm{lex}} (u_{i_1\dots i_r}, y_{i_1\dots i_r}) \in \mathscr{D}_{int}\Big] \,\textrm{ and} &
    \\
       \,\,\,\Big[(e_{n}, f_{n}, u_n, y_n) \in \mathscr{D}_{\omega_n}   \forall n\in\{1,\dots N\}\Big] & \textrm{and} 
    \\
      \,\,\,\Big[(e_{i_1\dots i_r}, f_{i_1\dots i_r},u_{i_1\dots i_r}, y_{i_1\dots i_r}) \in \mathscr{D}_{i_1\dots i_r} &  \forall\,i\in\bigcup\limits_{r=2}^N \I{r} \Big]  
  \end{array}\right\} \,.
\end{equation}
Substituting the defining relations of Dirac structures $\mathscr{D}_{\omega_n}$ and $\mathscr{D}_{i_1\dots i_r}$ into the interconnection Dirac structure $\mathscr{D}_{int}$ eliminates all auxiliary variables and reproduces the defining relations of \eqref{hiresDleaner}. 
Conversely, every element of $\mathscr{D}_N$
determines unique variables satisfying the defining relations of the constituent Dirac structures $\mathscr{D}_{\omega_n}$, $\mathscr{D}_{i_1\dots i_r}$ and $\mathscr{D}_{int}$. Hence the two constructions define the same subbundle. Diagrammatically, the refined equations of motion are represented by  
\begin{equation*}\label{fig:highesteom}
    \centering
\includegraphics[width=1.0\linewidth]{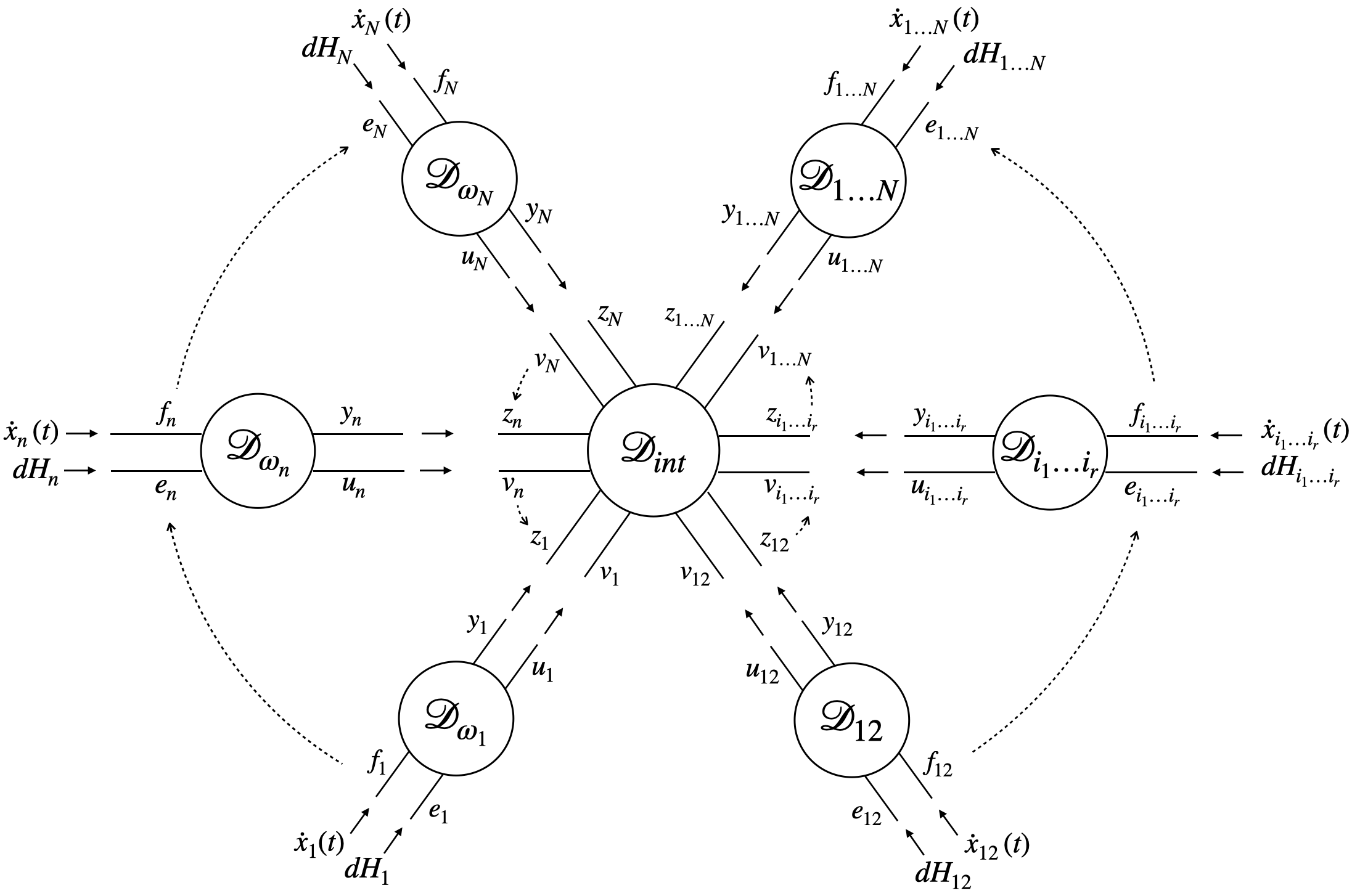}
\end{equation*} \enlargethispage{12pt}
From the perspective of Dirac geometry, the construction above is an instance of the composition (or interconnection) of Dirac structures through shared port variables. The interested reader is referred to~\cite{vanderschaftintegrab} for the general definition of the composition of Dirac structures.
\newpage
\section{Conclusions}
\noindent The results obtained in this work establish a precise operational dictionary between a single classical canonical Hamiltonian system and its high-resolution port-Hamiltonian formulation. 
Rather than viewing port-Hamiltonian theory as merely containing Hamiltonian mechanics at its lowest resolution, we have demonstrated that highly resolved port-Hamiltonian descriptions are already encoded in a classical canonical Hamiltonian system through an interplay of  the configuration space topology and the complexity of the specified dynamics. 

Technically, we devised a differential topological method to pass from a canonical Hamiltonian to a port-Hamiltonian description which
is conceptually analogous to a Legendre transformation that passes to a Lagrangian description. 
Moreover, we showed that the total port space is technically and conceptually most insightfully described by the bundle sum $\solder$ that interpolates between the ordinary product $\times$ of bundles with arbitrary base spaces and the Whitney sum $\oplus$ of bundles with the same base spaces. We showed that an essential part of the defining conditions for the Dirac structure is automatically induced from this way of merging port spaces to a total port space, and thus does not need to be supplied as a modelling choice. Finally, we demonstrated how one can isolate the purely topological part of the Dirac equation.

Crucially, this changes how we can read a physical system. The level of resolution of internal structure is no longer an arbitrary modelling choice in port-Hamiltonian theory, but a precise mathematical consequence governed by the topological structure of the configuration space, the genericity of the Hamiltonian function and a choice of how much one wishes to include or ignore of the configuration space topology.
Thus the framework we devised opens the door to a systematic, geometric exploration of internal substructures in Hamiltonian physical systems that were always present, but previously invisible through the lens of the Hamiltonian description alone.

 In order to properly place the implications of our general result in context and indicate how it can be used to extend the already very broad range of application of port-Hamiltonian theory, it is worth briefly outlining the scope of the framework beyond the canonical setting considered here. First of all, Dirac structures, unlike symplectic structures, exist on smooth manifolds of arbitrary dimension and thus also accommodate Poisson systems~\cite{courant1990dirac}. Singular Lagrangian systems, for which the Legendre transform is not invertible, can then be described by generalised energy storages in terms of Lagrangian submanifolds rather than a Hamiltonian function~\cite{vanderschaft2020dirac, krhac2026generalised, barbero2019dirac, beattie2019linear, vanderschaft2013dae}. More broadly, Dirac geometry already provides the foundation for geometric formulations outside the strict port-Hamiltonian setting, including the implicit Lagrange--Dirac formalism on the Pontryagin bundle~\cite{yoshimura2006dirac1, yoshimura2006dirac2} and Dirac-geometric treatments of non-holonomic systems~\cite{barbero2019dirac, de2019dirac}. Physically, the port-Hamiltonian framework is by no means restricted to conservative mechanical systems. Instead, it provides a unified modelling paradigm for multi-physics domains—including electrical circuits, mechanical networks, electromechanical transducers, and hydraulic systems \cite{duindam2009, vanderschaft2014introductory}. While dissipation is readily incorporated through resistive ports~\cite{vanderschaft2014introductory, maschke1992port}, the framework further extends to irreversible, non-isothermal phenomena through thermodynamic port-Hamiltonian formulations that explicitly satisfy the fundamental laws of thermodynamics~\cite{ramirez2013, ramirez2016}. Finally, beyond finite-dimensional systems, port-Hamiltonian theory has also developed into a rich theory for distributed-parameter systems governed by partial differential equations. One line of research extends the geometric foundations of the theory to infinite-dimensional settings through Stokes--Dirac structures and related decomposition frameworks~\cite{vanderschaft2002hamiltonian, brugnoli2023stokes, rashad2025continuum, rashad2021ideal, califano2021geometric}. A complementary operator-theoretic approach formulates port-Hamiltonian systems as boundary control systems and abstract Cauchy problems, with an emphasis on well-posedness, semigroup theory, and control~\cite{jacob2012linear, le-gorrec2005}.

Remarkably, the techniques and results developed in this article contribute to a further extension beyond this already very broad spectrum covered by existing port-Hamiltonian theory. This is because they open up the arena for a non-trivial port-formulation of classical stochastic processes beyond those driven by a Hamiltonian or diffusion. A possible extension to the Wigner formulation of quantum dynamics is thus also in reach.  

\begin{center}
   {\bf\small ACKNOWLEDGMENTS}
\end{center}
\noindent The authors acknowledge funding from the European
Union Horizon Europe MSCA Grant No. 101073558 (ModConFlex).

\bibliography{biblio}

\end{document}